\documentclass[pra,aps,showpacs,groupedaddress,superscriptaddress,twocolumn]{revtex4-1}

\usepackage[utf8x]{inputenc}
\usepackage{color}
\usepackage{bbm}
\usepackage{amsfonts,amsmath,amssymb,stmaryrd}
\usepackage{graphicx}
\usepackage{subfigure}  
\usepackage{bbm}
\usepackage{hyperref}
\usepackage{epsfig}
\usepackage{mathrsfs}
\usepackage{verbatim}
\usepackage{centernot}
\usepackage{ulem}

\renewcommand{\l}{\left(}
\renewcommand{\r}{\right)}

\newcommand{\bra}[1]{\langle#1|}
\newcommand{\ket}[1]{|#1\rangle}

\renewcommand{\H}{\hat{\mathcal{H}}}

\renewcommand{\a}{\hat{a}}

\newcommand{\ad}{\hat{a}^\dagger}

\newcommand{\G}{\hat{\Gamma}}

\newcommand{\MF}{\text{MF}}

\newcommand{\RG}{\text{RG}}

\newcommand{\ph}{\text{ph}}
\newcommand{\IB}{\text{IB}}
\newcommand{\B}{\text{B}}

\usepackage{array}

\usepackage{cancel,ifthen}
\newcommand{\cmnt}[2][NoInPuT]{\ifthenelse{\equal{#1}{NoInPuT}}{}{{\color{red}\sout{#1}}} {\color{blue} #2}}

\usepackage{bm}	
\renewcommand{\vec}[1]{\bm{#1}}

\begin{document}
\normalem	

\title{Bose polarons in ultracold atoms in one dimension: beyond the Fr\"ohlich paradigm}

\author{Fabian Grusdt}
\affiliation{Department of Physics, Harvard University, Cambridge, Massachusetts 02138, USA}

\author{Gregory E. Astrakharchik}
\affiliation{Departament de Fisica, Campus Nord B4-B5, Universitat Politecnica de Catalunya, E-08034 Barcelona, Spain}

\author{Eugene Demler}
\affiliation{Department of Physics, Harvard University, Cambridge, Massachusetts 02138, USA}

\date{\today}

\begin{abstract}
Mobile impurity atoms immersed in Bose-Einstein condensates provide a new platform for exploring Bose polarons. Recent experimental advances in the field of ultracold atoms make it possible to realize such systems with highly tunable microscopic parameters and to explore equilibrium and dynamical properties of polarons using a rich toolbox of atomic physics. In this paper we present a detailed theoretical analysis of Bose polarons in one dimensional systems of ultracold atoms. By combining a non-perturbative renormalization group approach with numerically exact diffusion Monte Carlo calculations we obtain not only detailed numerical results over a broad range of parameters but also qualitative understanding of different regimes of the system. We find that an accurate description of Bose polarons requires the inclusion of two-phonon scattering terms which go beyond the commonly used Fr\"ohlich model. Furthermore we show that when the Bose gas is in the strongly interacting regime, one needs to include interactions between the phonon modes. We use several theoretical approaches to calculate the polaron energy and its effective mass. The former can be measured using radio-frequency spectroscopy and the latter can be studied experimentally using impurity oscillations in a harmonic trapping potential. We compare our theoretical results for the effective mass to the experiments by Catani et al. [PRA 85, 023623 (2012)]. In the weak-to-intermediate coupling regimes we obtain excellent quantitative agreement between theory and experiment, without any free fitting parameter. We supplement our analysis by full dynamical simulations of polaron oscillations in a shallow trapping potential. We also use our renormalization group approach to analyze the full phase diagram and identify regions that support repulsive and attractive polarons, as well as multi-particle bound states.
\end{abstract}

\maketitle

\section{Introduction}
When a mobile particle interacts with a surrounding bath of bosons, it becomes dressed by a cloud of excitations and forms a polaron~\cite{Landau1948,Mahan2000}. As a result many of its properties, like the effective mass, are strongly modified compared to those of the bare particle. Impurity atoms immersed in a Bose gas provide a promising new platform for studying the long standing polaron problem. Advantages of such systems include the tunability of both interactions~\cite{Chin2010,Olshanii1998,Tempere2009} and the single particle dispersion~\cite{Bloch2008}. For example, both impurity and host atoms can be realized in a quasi one-dimensional (1D) geometry. This situation was realized experimentally by the Florence group~\cite{catani2012quantum} and will be considered throughout this paper. Recent experiments also demonstrated the existence of strongly coupled Bose polarons in one \cite{catani2012quantum} and three dimensional systems~\cite{Hu2016PRL,Jorgensen2016PRL}.

\begin{figure*}[t]
\centering
\epsfig{file=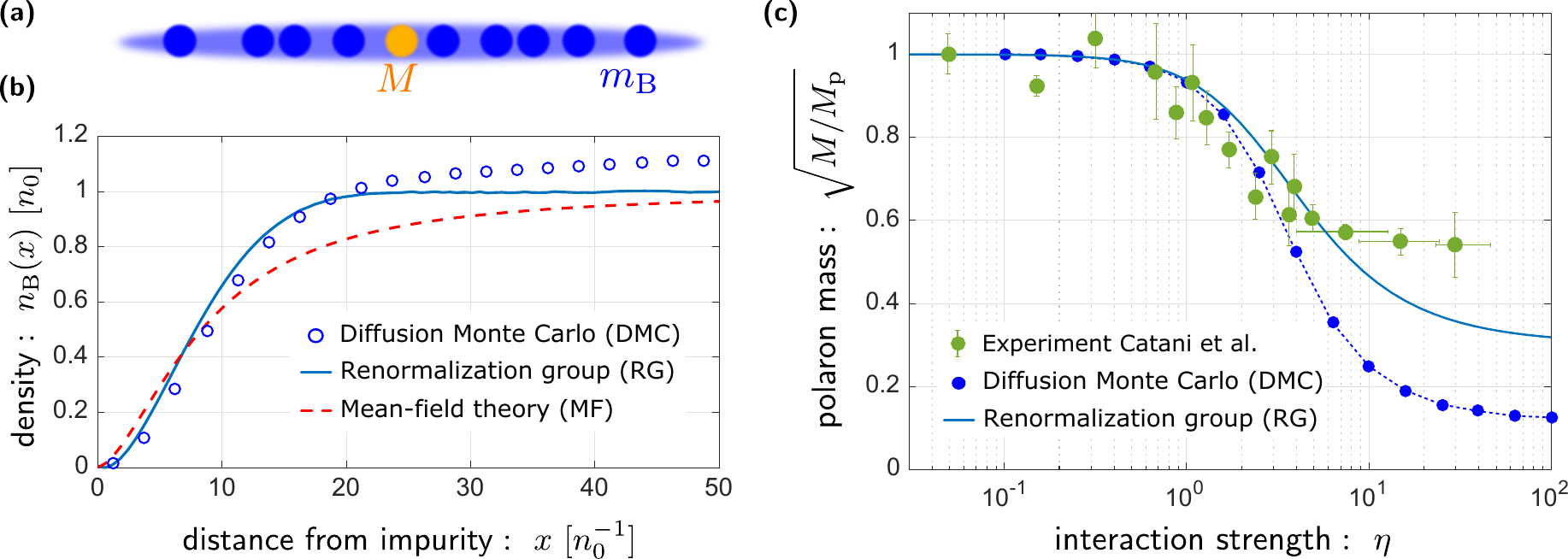, width=0.9\textwidth}
\caption{Strongly coupled polarons in one dimension: (a) System sketch: we consider the problem of a mobile impurity interacting with a quasi-one dimensional Bose gas. We perform calculations based on various theoretical methods, including a RG approach and diffusion Monte Carlo (DMC) calculations. (b) Deep in the Bogoliubov regime the RG method yields accurate results for the density profile of bosons around the impurity (i.e. the impurity-boson correlation function) even for strong impurity-boson interactions. Importantly the RG improves predictions by simpler mean-field (MF) calculations based on an ansatz of uncorrelated phonons. Here we used a ratio of impurity-boson to boson-boson interaction strengths of $\eta=1000$ (very strong repulsion) and considered the case of high density with $n_0|a_{\rm BB}| = 144$, where $n_0$ is the density of the Bose gas and $a_{\rm BB}$ the one-dimensional boson-boson scattering length. (c) We also calculate the effective polaron mass $M_{\rm p}$ and compare our results to experimental measurements by Catani et al.~\cite{catani2012quantum}. In the weak-to-intermediate coupling regimes we obtain excellent quantitative agreement between theory and experiment, without any free fitting parameters.}
\label{fig:Intro}
\end{figure*}

Numerous theoretical works have addressed the problem of a mobile impurity in an ultracold quantum gas, see Refs.~\cite{Devreese2013,Grusdt2015Varenna} for reviews. However, they were either based on an effective Fr\"ohlich Hamiltonian to describe the polaron \cite{Bruderer2007,BeiBing2009,Tempere2009,Casteels2012,Kain2014,Grusdt2015RG,Shchadilova2016,Vlietinck2015,Kain2016} or used truncated wave functions with only a few excitations \cite{Li2014,Levinsen2015}. Notable exceptions include a third-order perturbative treatment of the problem \cite{Christensen2015}, a self-consistent T-matrix calculation \cite{Rath2013}, a mean-field (MF) analysis beyond the Fr\"ohlich Hamiltonian \cite{Shchadilova2016PRL}, diffusion Monte Carlo calculations based on the full microscopic Hamiltonian \cite{Ardila2015,Parisi2016} and approximate analytical descriptions \cite{Dehkharghani2015,Volosniev2017}. Recently Virial expansion techniques have also been used to study spectra of Bose polarons \cite{Sun2017} and a flow-equation approach has been applied to the problem \cite{Volosniev2017}.

A number of important questions remain open and a complete theoretical understanding of Bose polarons at arbitrary couplings is lacking. Most strikingly, the phase diagram in the strongly interacting regime is still a subject of debate. In this paper we focus on a system where both the impurity and the Bose gas are constrained to one dimension. When the impurity is interacting with only a single boson, a two-particle bound state exists already for infinitesimal attractive interactions. If the mass of the impurity is infinite and multiple bosons without mutual interactions are considered, this gives rise to an infinite series of multi-particle bound states. The fate of these many-body eigenstates in a regime where the impurity is mobile and the Bose gas is interacting, is unclear. A related question concerns the regimes of validity of different effective polaron models, including the celebrated Fr\"ohlich Hamiltonian.

In two and three dimensional systems, the MF approach \cite{Shchadilova2016PRL} is a convenient theoretical tool that can be used to study models beyond the simplified Fr\"ohlich Hamiltonian. It is a non-perturbative method which includes strong correlations between the phonons and the impurity, whereas phonon-phonon correlations are neglected. For example, one can include two phonon scattering terms that are crucial for the accurate description of few-body aspects of the system including the existence of bound states between the impurity and host bosons \cite{Ardila2015,Shchadilova2016PRL}. The spectral function of the impurity in three dimensional systems obtained using the MF approach~\cite{Shchadilova2016PRL} was in good agreement with experimental results~\cite{Hu2016PRL,Jorgensen2016PRL}. The applicability of the MF approach to 1D systems has not been clarified yet. One of the indications that 1D systems are special is the unphysical logarithmic infrared divergence of the polaron energy, which is absent in higher dimensions. The physical origin of this divergence is the enhanced role of quantum fluctuations in 1D systems. These are essentially the same fluctuations that are responsible for the absence of true Bose-Einstein condensation in homogeneous 1D systems even at zero temperature~\cite{Hohenberg1967,Mermin1966}.

Theoretical issues raised above provide a considerable challenge for quantitative analysis of the experiment~\cite{catani2012quantum}, where 1D Bose polarons have been realized at strong couplings for the first time. In fact, even in the weak coupling regime, the agreement of the measured effective polaron mass with earlier theoretical calculations based on the effective Fr\"ohlich Hamiltonian has not been satisfactory. In order to obtain quantitative agreement, the impurity-boson coupling had to be multiplied by a factor of $3.15$ in Ref.~\cite{catani2012quantum}. Moreover, at stronger couplings a saturation of the effective mass has been observed~\cite{catani2012quantum}, which lacked theoretical explanation so far.

Most of the earlier theoretical work focused on equilibrium properties of polarons. The experiments with polarons that have been carried out in 1D quantum gases so far, including measurements by Catani et al.~\cite{catani2012quantum}, all probed non-equilibrium impurity dynamics \cite{Fukuhara2013,Scelle2013}. Thus theoretical analysis has to study not only strongly interacting systems, but also understand its dynamical properties and their connection to equilibrium quantities. This provides an additional challenge, since most of the standard tools, such as Monte-Carlo methods, are not applicable. First steps in this direction have been taken in Refs.~\cite{Zvonarev2007,Bonart2012,Bonart2013,Volosniev2015,Lampo2017}. 

In this paper we address the questions raised above and provide a detailed theoretical analysis of the Bose polaron problem in one dimension. We consider a mobile impurity of mass $M$ interacting with a 1D Bose gas, see Fig.~\ref{fig:Intro}(a). We then compare our theoretical methods, which leads us to an understanding of which terms in the microscopic Hamiltonian contribute most to the polaron properties. As an example, in Fig.~\ref{fig:Intro}(b) we show that the depletion of the Bose gas around the impurity can be described accurately by a semi-analytical renormalization-group (RG) approach \cite{Grusdt2015RG,Grusdt2016RG,Grusdt2016RGBEC} when the Bose gas is deep in the Bogoliubov regime. Moreover, we use our theoretical methods to analyze the experiment by Catani et al.~\cite{catani2012quantum} in detail. In particular we calculate the effective polaron mass. In the weak-to-intermediate coupling regimes we obtain excellent agreement with the experimental data, see Fig.~\ref{fig:Intro} (c). Our results moreover provide an important test case for theories of Bose polarons at strong couplings, applicable also in higher dimensions~\cite{Grusdt2016RGBEC}. 

A special feature of our work is the comparison of analytical analysis with numerical calculations starting from the full microscopic Hamiltonian and based on the diffusion Monte Carlo (DMC) method \cite{Anderson1975,Boronat1994,Matveeva2016}, supplemented by variational Monte Carlo (VMC) calculations. In addition we present results from time-dependent MF simulations, following Refs.~\cite{Shashi2014RF,Shchadilova2016PRL}, to study impurity dynamics.

Our paper is organized as follows. After briefly summarizing our main results in the following section, we introduce the model in Sec.~\ref{sec:Model}. In Sec.~\ref{eq:FroehlichLimit} we consider the weak coupling limit and compare our calculations to the experimental data from Ref.~\cite{catani2012quantum}. We discuss the effect of two-phonon terms in Sec.~\ref{sec:TwoPhononTerms}. A detailed RG analysis is presented in Appendix \ref{appdx:RGapproach}, from which we derive the polaron phase diagram. In Sec.~\ref{sec:PolaronDynamics} a time-dependent MF theory is applied to analyze polaron dynamics as in the experiment by Catani et al. Section~\ref{sec:BeyondBogoliubov} is devoted to a discussion of phonon-phonon interactions in the polaron cloud. The variational and diffusion Monte Carlo methods are presented in Sec.~\ref{subsec:DMCapproach}. We close with a summary and an outlook in Sec.~\ref{sec:SummOut}.

\section{Summary of results}
\label{subsec:SummaryRes}

\begin{figure}[t!]
\centering
\epsfig{file=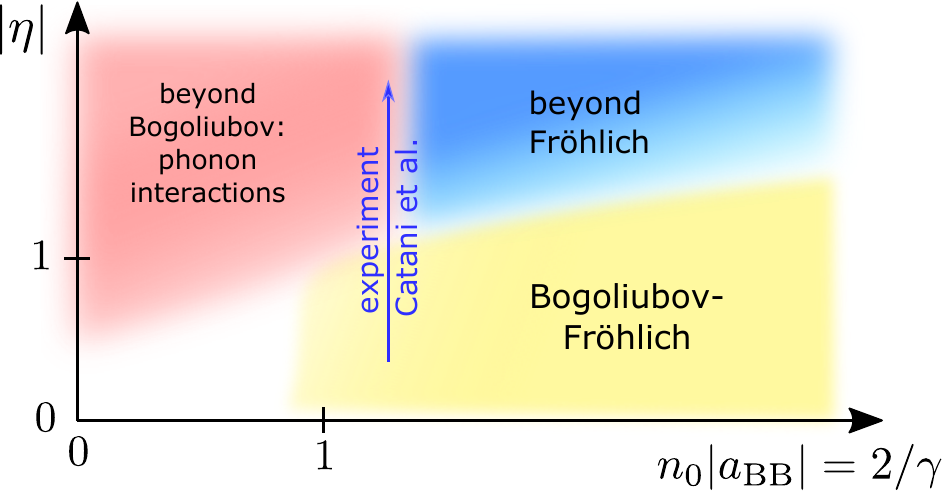, width=0.42\textwidth}
\caption{A diagram with different physical regimes of Bose polarons: For weak impurity-boson interactions $\eta$ an impurity in a Bose gas can be described by the Fr\"ohlich Hamiltonian. For strong interactions, two-phonon terms beyond the Fr\"ohlich model
have to be included. For strongly interacting bosons, $\gamma  \gtrsim 1$ where $\gamma = 2 / n_0 |a_{\rm BB}|$ is the dimensionless interaction strength of the Bose gas \cite{Bloch2008}, the Bogoliubov approximation breaks down and phonon-phonon interactions also have to be included. For weakly interacting bosons, $\gamma \ll 1$, we identify a regime where the Bogoliubov approximation for the Bose gas is justified (beyond Fr\"ohlich regime).}
\label{fig:Intro2}
\end{figure}

Bose polarons have been commonly investigated theoretically using an effective Fr\"ohlich Hamiltonian $\H_{\rm F}$ \cite{Tempere2009,catani2012quantum,Casteels2012}. This is justified for weak interactions between the impurity and bath particles, see Fig.~\ref{fig:Intro2}. In the present work we show that in this regime, the effective Fr\"ohlich model describes accurately the experimental results for the effective mass~\cite{catani2012quantum}, without any free fitting parameter. We point out the importance of high-energy phonons (at momenta of the order of the inverse healing length $\xi$), which have not been treated accurately in the previous analysis of the experimental data~\cite{catani2012quantum,Bonart2012,Bonart2013}. When the Bose gas is weakly interacting, we find a regime where perturbative treatments of impurity-phonon interactions \cite{Ardila2015,Parisi2016} fail. The Fr\"ohlich model is still valid for these parameters however.

For stronger impurity-boson interactions, the Fr\"ohlich Hamiltonian is no longer sufficient and two-phonon processes $\H_{\rm 2 ph}$ (see Eq.~\eqref{eq:H2phDef} for details) have to be included to describe the depletion of the condensate correctly~\cite{Rath2013}, allowing also for molecular states~\cite{Rath2013,Shchadilova2016PRL,Grusdt2016RGBEC}. To solve the extended polaron Hamiltonian, we first use MF theory in the spirit of Ref.~\cite{Shchadilova2016PRL} and show that it predicts a logarithmic divergence of the polaron energy with the infrared momentum cut-off. The divergence can be regularized by a more accurate RG calculation, which we benchmark in 1D by comparing to our numerically exact DMC results. For large boson densities, corresponding to a regime where the Bogoliubov approximation can be used to describe the Bose gas, we find good agreement of the RG with exact DMC predictions, see Fig.~\ref{fig:Intro} (b).

Due to the presence of the two-phonon terms, the MF and RG approaches predict a saturation of the effective polaron mass to a large but constant value for strong interactions with the impurity, see Fig.~\ref{fig:Intro} (c). This effect can be understood as a shift of the position of the impurity-boson Feshbach resonance to the attractive side ($\eta <0$ in Fig.~\ref{fig:Intro} (c)) due to the interactions with the surrounding Bose gas. We note that at the Feshbach resonance the interaction strength diverges and the polaron mass becomes infinite \cite{Fuchs2005}. This can be understood by considering a finite-size system of $N$ bosons with hard-core interactions with the impurity. Because the bosons cannot penetrate the impurity, the latter can only move together with the bosons. Hence the polaron mass is proportional to $N$ and diverges in the thermodynamic limit.

Qualitatively, a saturation of the polaron mass at strong impurity-boson interactions has been observed experimentally~\cite{catani2012quantum} using breathing oscillations of the impurity in a harmonic trap. To analyze these measurements further, we perform full dynamical simulations of the polaron trajectory using time-dependent MF theory~\cite{Shashi2014RF,Shchadilova2016PRL}. Our calculations show that while the impurity oscillations provide a powerful tool to determine the effective polaron mass, the accuracy can be limited by the inhomogeneity of the Bose gas, in particular for strong interactions. 

When the boson density $n_0$ becomes small, the Bogoliubov theory of the interacting Bose gas breaks down. In this regime phonon-phonon interactions described by $\H_{\rm ph-ph}$ (see Eq.~\eqref{eq:defHphph} for details) have to be included for a valid description of the polaron, see Fig.~\ref{fig:Intro2}. From the comparison of our most reliable DMC calculations (including $\H_{\rm ph-ph}$) with RG predictions (without $\H_{\rm ph-ph}$) we find that phonon-phonon interactions always need to be included to obtain quantitative agreement for the polaron energy and mass.

From the comparison of our DMC and RG calculations, we also conclude that the experiment by Catani et al.~\cite{catani2012quantum} has been performed in a regime where all terms in the Hamiltonian ($\H_F$, $\H_{\rm 2ph}$ and $\H_{\rm ph-ph}$) are relevant. In particular the interactions between Bogoliubov phonons in the bath already play a role. Moreover our simulations of polaron dynamics suggest that the inhomogeneity of the Bose gas should be included when analyzing results of the experiment. We point out how  additional experiments can shed new light on the physics of strongly coupled Bose polarons in one dimension.

\section{Model}
\label{sec:Model}
Our starting point is a single impurity interacting with a Bose gas in one dimension, see Fig.~\ref{fig:Intro} (a). The bosons also have mutual interactions. This situation can be described by the following microscopic Hamiltonian,
\begin{multline}
 \H = \int dx ~ \hat{\phi}^{\dagger}(x) \left[ -\frac{\partial_x^2}{2 m_\B} + \frac{g_{\text{BB}}}{2} \hat{\phi}^\dagger(x) \hat{\phi}(x)\right] \hat{\phi}(x)\\
+ \int dx ~ \hat{\psi}^{\dagger}(x) \left[ -\frac{\partial_x^2}{2 M} + g_{\IB} \hat{\phi}^{\dagger}(x) \hat{\phi}(x) \right] \hat{\psi}(x),
 \label{eq:Hmicro}
\end{multline}
where $\hat{\phi}(x)$ stands for the Bose field operator, $\hat{\psi}(x)$ is the impurity field and $\hbar =1$. The boson (impurity) mass is $m_{\rm B}$ ($M$) and $g_{\rm BB}$ ($g_{\rm IB}$) denote the boson-boson and impurity-boson coupling constants respectively. We assume that only a single impurity is present in the homogeneous Bose gas with density $n_0$. Experimentally this corresponds to a situation with sufficiently low impurity concentration, ideally with less than one impurity per healing length $\xi$.

The ground state of this Hamiltonian can be calculated efficiently for up to $N \gtrsim 200$ bosons by means of the DMC method. We provide the technical details of this approach in Sec.~\ref{subsec:DMCapproach}, although we compare to DMC results in earlier sections.

\subsection{Polaron description}
\label{subsec:PolaronDescription}
To arrive at a polaron description of the impurity problem described above, we express the boson field operator $\hat{\phi}(x)$ in terms of Bogoliubov phonons $\a_k$ \cite{Mathey2004,Tempere2009,Devreese2013,Grusdt2015Varenna}. To this end we write
the Fourier components $\hat{\phi}_k = (2 \pi)^{-1/2} \int dx~ e^{i k x} \hat{\phi}(x)$ as
\begin{equation}
\hat{\phi}_k = \cosh \theta_{k} \a_{k} - \sinh \theta_{k} \ad_{-k},
\end{equation}
where $\theta_k$ is chosen as in the usual Bogoliubov theory for a weakly interacting Bose gas~\cite{Pitaevskii2003},
\begin{equation}
\begin{array}{c}
\cosh \theta_{k}  \\
\sinh \theta_{k}
\end{array}
= \frac{1}{\sqrt{2}} \l \frac{k^2/2 m_{\text{B}} + g_{\text{BB}} n_0 }{ \omega_k } \pm 1 \r^{1/2}.
\end{equation}
Here the Bogoliubov dispersion is given by
\begin{equation}
\omega_k = c k \l 1 + \frac{1}{2} k^2 \xi^2 \r^{-1/2},
\label{eq:omegakDef}
\end{equation}
where $c= \sqrt{g_{\text{BB}} n_0 / m_{\text{B}}}$ and $\xi = 1 / \sqrt{2 m_{\text{B}} g_{\text{BB}} n_0}$ are the speed of sound and the healing length in the limit of weak interactions.

So far we have only applied a basis transformation, allowing us to express the Bose field $\hat{\phi}(x)$ in terms of Bogoliubov phonons $\a_k$. Thereby the Hamiltonian in Eq.~\eqref{eq:Hmicro} can be written as
\begin{equation}
\H = \H_F + \H_{\rm 2ph} + \H_{\rm ph-ph}
\label{eq:Hfull}
\end{equation}
without any approximation. The first term corresponds to the effective Fr\"ohlich Hamiltonian,
\begin{multline}
\H_F = \int dk ~ \omega_k \ad_k \a_k + \frac{\hat{p}^2}{2 M} +\\
+ g_\IB n_0 + \int dk ~ V_k e^{i k \hat{x}} \l \ad_k + \a_{-k} \r.
\label{eq:HFdef}
\end{multline}
Here $\hat{p}$ and $\hat{x}$ denote the momentum and position operators of the impurity in first quantization. The scattering amplitude is given by~\cite{Tempere2009,Grusdt2015Varenna}
\begin{equation}
V_k =  \sqrt{n_0} (2 \pi)^{-1/2} g_\IB W_k, \quad W_k= \l \frac{(\xi k)^2}{2 + (\xi k)^2} \r^{1/4}.
\label{eq:VkDef}
\end{equation}
The second term, $\H_{\rm 2ph}$, describes two-phonon scattering processes~\cite{Rath2013,Shchadilova2016PRL} and reads
\begin{multline}
\H_{\rm 2ph} = \frac{g_\IB}{2 \pi} \int dk dk'~  \l \cosh \theta_k \ad_{k} - \sinh \theta_k \a_{-k} \r \\
\times  \l \cosh \theta_{k'} \a_{k'} - \sinh \theta_{k'} \ad_{-k'} \r ~ e^{i (k-k') \hat{x}}.
\label{eq:H2phDef}
\end{multline}

The two terms above, $\H_F$ and $\H_{\rm 2ph}$, provide an accurate model for the polaron problem when the Bose gas can be treated within Bogoliubov theory. This mean-field description of the interacting bosons assumes a macroscopic occupation of the condensate which is absent in one dimension \cite{Mermin1966,Hohenberg1967}. As pointed out by Lieb and Liniger~\cite{Lieb1963a,Lieb1963}, some quantities including the total energy can nevertheless be calculated accurately using Bogoliubov theory in the regime of weak interactions. This is the case when the dimensionless coupling strength $\gamma \lesssim 2$ is sufficiently small~\cite{Lieb1963a}, where
\begin{equation}
\gamma = \frac{2}{n_0 |a_{\rm BB}|} = \frac{m_{\rm B} g_{\rm BB}}{n_0}
\label{eq:defGamma}
\end{equation}
and the 1D boson-boson $s$-wave scattering length is given by the relation
\begin{equation}
a_{\rm BB}=-2/(m_{\rm B}g_{\rm BB}).
\label{eq:aBB}
\end{equation}

In contrast, for strong interactions $g_{\rm BB}$ or small densities $n_0$, when $\gamma \gtrsim 2$ is large, the Bogoliubov description of the Bose gas breaks down. In this case additional interactions between the Bogoliubov phonons $\a_k$ have to be included, which we summarize in $\H_{\rm ph-ph}$. These terms are of order
\begin{equation}
\H_{\rm ph-ph}=\mathcal{O} (\a_k^3) + \mathcal{O} (\a_k^4),
\label{eq:defHphph}
\end{equation}
but do not involve impurity operators. Since we do not need the expression for $\H_{\rm ph-ph}$ in terms of Bogoliubov phonons in the following, we will not write them out explicitly.

The terms in Hamiltonian \eqref{eq:Hfull} give rise to different physical regimes of the Bose polaron. For small $g_{\rm IB}$, the ground state  corresponds to a free impurity and phonon contributions are negligible. In this regime the polaron energy is determined by $g_{\rm IB} n_0$, sometimes referred to as the mean-field shift. For stronger couplings states with one phonon need to be included, which has been done systematically by using perturbation theory in Refs.~\cite{Ardila2015,Parisi2016}. This approach is valid when $\gamma \ll 1$ and
\begin{equation}
\eta \ll 2 \gamma^{3/4} \approx 3  ( n_0 |a_{\rm BB}|)^{-3/4}.
\label{eq:validityPerturbationTheory}
\end{equation}
Beyond this coupling strength multi-phonon states need to be added in the ground state wavefunction to describe correctly the polaron energy its effective mass \cite{Shashi2014RF,Grusdt2015RG}. To go beyond the Fr\"ohlich model, two-phonon terms need to be included in the Hamiltonian \cite{Rath2013,Shchadilova2016PRL}, see Fig.~\ref{fig:Intro2}. When the polaron cloud contains many phonons and boson-boson interactions become large, their interactions need to be taken into account as well. On a mean-field level this can be done by using the Gross-Pitaevskii equation, see Sec.~\ref{subsec:GPElimit} for a discussion.

\subsection{Experimental considerations}
\label{subsec:Experiments}
Quasi-1D Bose gases have been realized in several experiments (see Ref.~\cite{Cazalilla2011} for a review). Mobile impurities can be realized by using a second atomic level, as in the recent experimental observation of Bose polarons in a 3D Bose-Einstein condensate \cite{Jorgensen2016PRL}. Alternatively a second atomic species can be added, as has been demonstrated in Refs.~\cite{Spethmann2012,catani2012quantum,Hu2016PRL,Rentrop2016}.

In this paper we analyze the experimental results from Ref.~\cite{catani2012quantum}. We use the dimensionless parameter
\begin{equation}
\eta = g_{\rm IB} / g_{\rm BB}
\end{equation}
introduced therein to quantify the interaction strength between impurity and boson with respect to the fixed strength of interactions in the bath. The mass ratio in the experiment was $M/m_{\rm B} = 41/87 = 0.47$ and for $n_0$ we use the peak density of the Bose gas, which was estimated to be $n_0= 7 / \mu {\rm m}$ in Ref.~\cite{catani2012quantum}. The coupling constant in the bath is taken equal to $g_{\rm BB} = 2.36 \times 10^{-37} \rm J m$ as in Ref.~\cite{catani2012quantum}.

Using Bogoliubov theory this yields the following estimates for the healing length, $\xi=0.15 \mu \rm m$ and the speed of sound, $c=3.38 \rm mm/s$. Using $\xi$ as a unit of length yields $n_0 \xi = 1.05$. A characteristic energy can be defined by $c/\xi = 3.6 \times 2 \pi \rm kHz$. Another important length scale is the 1D boson-boson scattering length $a_{\rm BB} = - 3.6 \mu \rm m$, see Eq.~\eqref{eq:aBB}.
The speed of sound calculated from the Bethe ansatz \cite{Lieb1963a,Lieb1963} at $\gamma = 2/(n_0 a_{\rm BB}) = 0.438$, see Eq.~\eqref{eq:defGamma}, equals to $c=3.20 \rm mm/s$ so that the contribution of quantum fluctuations is about 5\%. In order to avoid a possible issue which value (Bogoliubov theory or Bethe ansatz) is used, we take the healing length $\xi$ as a unit of length for comparison with the MF theory and $a_{\rm BB}$ for the comparison with Monte Carlo results.

The temperature of the Bose gas $T = 350 (50) {\rm nK}$ measured in the experiment corresponds to $k_B T \approx 2 c/\xi$. The transverse confinement frequency of the bosons (the impurity) stated in Ref.~\cite{catani2012quantum} corresponds to $\omega_\perp \approx 9.4 ~ (12.5) c/\xi$ and justifies a description as a 1D system.

\begin{figure*}[t]
\centering
\epsfig{file=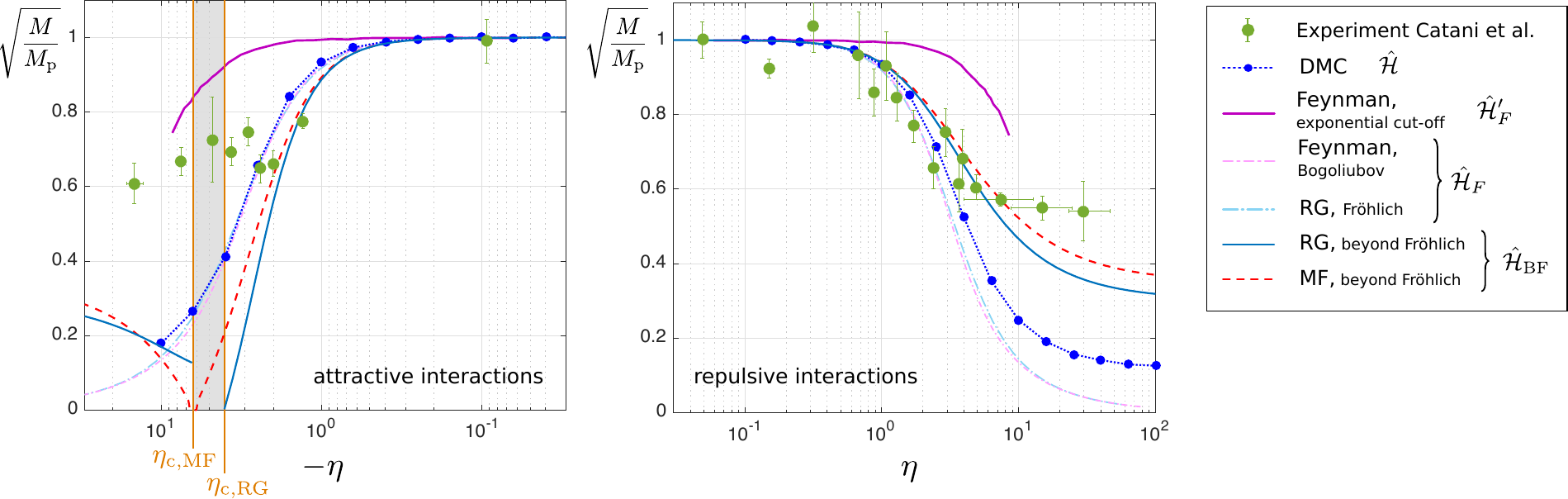, width=0.95\textwidth}
\caption{Effective polaron mass $M_{\rm p}$: We compare theoretical predictions for $M_{\rm p}$ with the experimental data by Catani et al.~\cite{catani2012quantum}, obtained from an indirect measurement of $\sqrt{M/M_{\rm p}}$ (for details see Ref.~\cite{catani2012quantum} and discussion below). The solid purple theoretical curve (Feynman, exponential cut-off) was also taken from Ref.~\cite{catani2012quantum} and corresponds to calculations based on Feynman's variational path integral approach, using the effective Fr\"ohlich Hamiltonian $\H_F'$ defined by Eqs.~\eqref{eq:wkPrime}, \eqref{eq:VkPrime}. Other theoretical curves are obtained from our own calculations, based on the Hamiltonians indicated in the legend. The shaded area on the attractive side corresponds to the regime where the RG (assuming the Hamiltonian $\H_{\rm BF} = \H_F + \H_{\rm 2ph}$) breaks down. DMC calculations were performed for $N=50$ particles and we checked that finite-size corrections are small.}
\label{fig:1DpolaronMass}
\end{figure*}

\section{Weak coupling limit: The Fr\"ohlich model}
\label{eq:FroehlichLimit}
We start by discussing the weak coupling limit where exact analytical results can be obtained within the Fr\"ohlich model. This enables a direct comparison between theory and experiment and moreover provides an important benchmark for subsequent analysis of the strong coupling regime.

Experimental studies by Catani et al.~\cite{catani2012quantum} included an indirect measurement of the effective polaron mass $M_{\rm p}$ for various values of the interaction strength $\eta$. The latter was tuned using a Feshbach resonance. The value of $M_{\rm p}$ was extracted from observations of breathing oscillations of the impurity interacting with the trapped Bose gas. The amplitude $\sigma$ of such oscillations is renormalized by a factor of $\sqrt{M/M_{\rm p}}$. This can be understood by considering an initially localized cloud of impurity atoms as in the experiment of Ref.~\cite{catani2012quantum}, with average kinetic energy $\sqrt{\langle \hat{p}^2 \rangle}/ 2M$. When the impurities are released their momentum distribution is adiabatically mapped to an identical distribution of polaron momenta. The resulting kinetic energy $\sqrt{\langle \hat{p}^2 \rangle}/ 2M_{\rm p}$ of polarons is subsequently converted into  potential energy $M \Omega_{\rm I}^2 \sigma^2 /2$ of the expanded atoms in a harmonic potential with trapping frequency $\Omega_{\rm I}$. The amplitude $\sigma$ thus provides a measure of the polaron mass, $\sigma = \sigma_0 \sqrt{M/M_{\rm p}}$ where $\sigma_0$ corresponds to non-interacting impurities. For a more detailed discussion see Ref.~\cite{catani2012quantum} and Sec.~\ref{sec:PolaronDynamics} below.

In  Fig.~\ref{fig:1DpolaronMass} we compare experimental results for the effective polaron mass $M_{\rm p}$ with predictions of our own numerical calculations. Before presenting a detailed technical analysis, we note that our data from four different theoretical methods show good agreement with the experimental results in the weak-to-intermediate coupling regimes. This is true for both repulsive and attractive interactions, without any free fitting parameter.

In their original analysis of the experiment, Catani and co-workers \cite{catani2012quantum} performed numerical calculations starting from an effective Fr\"ohlich Hamiltonian $\H_F'$ (defined in Sec.~\ref{subsec:UVfrohlich}) which they derived using bosonization techniques~\cite{Giamarchi2003}. Then they applied Feynman's variational path integral method~\cite{Feynman1955} to obtain the effective polaron mass within this model. Surprisingly, their prediction (thick purple line in Fig.~\ref{fig:1DpolaronMass}) showed appreciable disagreement with experimental results. Thus, either the evaluation of the polaron mass within the effective model was inaccurate, or the effective model itself is insufficient. To clarify, we provide the answers to the following questions before explaining them in detail below.

\begin{itemize}
\item[(i)] It has been suggested earlier that Feynman's variational path integral method becomes inadequate for intermediate coupling strengths $\eta$ if the impurity is light~\cite{Vlietinck2015,Grusdt2015RG,Shchadilova2016}. Is this method accurate for solving the effective Fr\"ohlich model $\H_F'$ for the experimental parameters? -- \emph{We will show that Feynman's approach is accurate in the regime considered here.}

\item[(ii)] To calculate the polaronic mass shift from Feynman's approach applied to the effective Fr\"ohlich model $\H_F'$, Catani et al.~\cite{catani2012quantum} used an estimate based on variational parameters instead of a full evaluation of the Green's function. This approximation can lead to sizable errors for polarons in ultracold quantum gases~\cite{Casteels2012,Grusdt2015RG}. Does the estimate work for the Fr\"ohlich model $\H_F'$ for the experimentally relevant parameters? -- \emph{The estimate leads to sizable deviations from the value of the effective polaron mass expected from the Fr\"ohlich Hamiltonian $\H_F'$. Nevertheless this effect is not sufficient to fully explain the disagreement with the experiment.}

\item[(iii)] The Fr\"ohlich model is only applicable for sufficiently weak impurity-boson interactions.
Where does this approximation break down? -- \emph{For $|\eta| \lesssim 3$, given the parameters form the experiment \cite{catani2012quantum}, a description in terms of a Fr\"ohlich Hamiltonian leads to quantitatively correct description.}

\item[(iv)] The effective Fr\"ohlich Hamiltonian $\H_F'$ derived from bosonization differs from $\H_F$ obtained from Bogoliubov theory in the way how the UV regularization is performed. Does this explain the disagreement between theory and experiment? -- \emph{In the 1D polaron cloud, phonons from all energies contribute to polaronic dressing and have strong effect on the polaron energy, effective mass, and dynamics. Therefore the polaron properties are sensitive to the details of the UV regularization. This causes the large deviations between earlier theoretical calculations based on $\H_F'$ and the experiment shown in Fig.~\ref{fig:1DpolaronMass}.}
\end{itemize}

\subsection{UV regularization of the Fr\"ohlich model}
\label{subsec:UVfrohlich}
The effective Fr\"ohlich Hamiltonian derived from Bogoliubov theory, see Eqs.~\eqref{eq:omegakDef}, \eqref{eq:HFdef} and \eqref{eq:VkDef}, does not require any UV regularization in one dimension and we can simply set the UV cut-off $\Lambda_0 = \infty$. At momenta $k \gtrsim 1/\xi$ the Bogoliubov dispersion becomes quadratic, $\omega_k \simeq k^2 / 2 m_{\rm B}$, making all quantities (the effective mass, the polaron energy, etc.) well behaved.

On the other hand, bosonization can be used to derive an effective Fr\"ohlich Hamiltonian \cite{catani2012quantum} for the impurity. It describes the interaction of the impurity with the Luttinger liquid formed by the 1D Bose gas \cite{Giamarchi2003,Cazalilla2011}. At long wavelengths, or low energies, the resulting Fr\"ohlich Hamiltonian $\H_F'$ has the same asymptotic form as $\H_F$ in Eq.~\eqref{eq:HFdef} which was derived from Bogoliubov theory. This provides additional evidence that the use of the Bogoliubov theory for treating the 1D Bose gas is justified.

The Fr\"ohlich Hamiltonian $\H_F'$ is obtained from Eq.~\eqref{eq:HFdef} by replacing $\omega_k$ and $V_k$ with different expressions $\omega_k'$ and $V_k'$, respectively. The bosonization approach relies on a linear phonon dispersion,
\begin{equation}
\omega_k' = c k,
\label{eq:wkPrime}
\end{equation}
and determines the scattering amplitude $V_k'$ at small $k$ (long wavelengths). Its ultraviolet (UV) cut-off, required for regularization of the model, is commonly represented by an exponential decay at a characteristic scale $k_c$ \cite{catani2012quantum} which is usually taken to be $k_c \sim 1/\xi$,
\begin{equation}
V_k' = \frac{g_{\rm IB}}{2 \pi} \sqrt{K |k|} e^{- k / 2 k_c}.
\label{eq:VkPrime}
\end{equation}
The dimensionless Luttinger parameter $K$ can be determined from Bethe-ansatz calculations. It is given by the ratio of the Fermi velocity $v_{\rm F} = \pi n_0 / m_{\rm B}$ and the speed of sound $c$,
\begin{equation}
K = \frac{v_F}{c} =\frac{\pi n_0}{c m_{\rm B}},
\label{Luttinger parameter}
\end{equation}
By using Eq.~(\ref{Luttinger parameter}) one confirms that the asymptotic behaviors of $V_k$ ($\omega_k$) and $V_k'$ ($\omega_k'$) in the infrared (IR) limit $k \ll 1/\xi$ are identical. For weakly interacting bosons, where the gas parameter $\gamma \ll 1$, see Eq.~\eqref{eq:defGamma}, is small, Bogoliubov theory provides an accurate value for the speed of sound as $c = \sqrt{g_{\rm BB} n_0 / m_{\rm B}}$. For a strongly interacting bath (Tonks-Girardeau regime), the speed of sound equals to the Fermi velocity, $c=v_F$. For intermediate interactions, Bethe ansatz can be used to obtain the speed of sound $c$ and Luttinger parameter $K$.

The main difference between the bosonization approach and Bogoliubov theory is in the treatment of phonons at high energies. In many transport and dynamic phenomena in 1D systems the main contributions come from the longest wavelength excitations \cite{Giamarchi2003} and the results are insensitive to the UV cut-off. In this spirit, Ref.~\cite{catani2012quantum} focused on low-energy phonons  and introduced the exponential cut-off at $k_c$ in Eq.~\eqref{eq:VkPrime} by hand. In the Bogoliubov theory, on the other hand, the phonon dispersion $\omega_k$ becomes non-linear for $k \gtrsim 1/\xi$, providing a natural UV cut-off scale in the model.

To obtain reasonable results from the effective Fr\"ohlich Hamiltonian $\H_F'$, the momentum cut-off $k_c$ has to be of the order $1/\xi$. Even for this choice the properties of UV phonons at momenta around $k \sim 1/\xi$ differ in the two models $\H_F$ and $\H_F'$. As has been shown by a dimensional analysis in Refs.~\cite{Grusdt2015RG,Grusdt2016RGBEC}, the high-energy modes with $k \sim 1/\xi$ play a crucial role in determining properties of the Bose polaron. Therefore we expect that there can be sizable quantitative differences between predictions by the two polaron Hamiltonians.

In Fig.~\ref{fig:1DpolaronMass} we calculate the polaron mass $M_{\rm p}$ starting from the Fr\"ohlich model $\H_F$ and using Feynman's variational path integral formalism \cite{Feynman1955,Tempere2009,Casteels2012} (purple dashed-dotted line in the figure). The same calculation was performed in Ref.~\cite{catani2012quantum}, but using the model $\H_F'$ with an exponential UV cut-off (purple solid line in the figure). Surprisingly, the Bogoliubov theory predicts a polaron mass which is about a factor of two larger than expected from assuming an exponential UV cut-off. This also explains the large deviations between the experimental data and theory based on $\H_F'$.

\subsection{Validity of the Fr\"ohlich model}
\label{subsec:validityFrohlich}
The Fr\"ohlich model is only valid when two-phonon processes can be ignored. This is the case when the depletion of the quasi-1D condensate is small, justifying also the assumption that phonon-phonon interactions cannot modify the polaron cloud substantially in this regime.

To derive an estimate when the Fr\"ohlich model is accurate, we apply standard MF theory~\cite{Lee1953,Devreese2013,BeiBing2009} (see also Sec.~\ref{subsec:MFtwoPhonon}) to the combined Hamiltonian $\H_F + \H_{\rm 2ph}$. In Appendix~\ref{sec:AppdxMF} we derive for the phonon number in the polaron cloud,
\begin{equation}
N_{\rm ph}^\MF =  \beta_\MF^2 \int dk~ \l \frac{V_k}{\omega_k + k^2/2M} \r^2.
\label{eq:NphMF}
\end{equation}
Here we assumed that the total momentum carried by the polaron vanishes, $p=0$. The result in Eq.~\eqref{eq:NphMF} is similar to the expression from the Fr\"ohlich model, which is obtained by setting $\beta_{\rm MF}=1$. When two-phonon terms are included, we obtain
\begin{equation}
\beta_\MF = \left[ 1 + \frac{g_\IB}{2 \pi} \int dk ~ \frac{W_k^2}{\omega_k + k^2/ 2 M} \right]^{-1}.
\label{eq:betaPzero}
\end{equation}
Therefore the depletion of the quasi-1D condensate can be described accurately by the Fr\"ohlich Hamiltonian provided that $\beta_\MF \ll 1$.

This yields the condition
\begin{equation}
|g_{\rm IB}| \ll g_{\rm IB}^c = 2 \pi \left[  \int dk ~ \frac{W_k^2}{\omega_k + k^2/2M} \right]^{-1}
\end{equation}
for the validity of $\H_F$. The integral can be estimated as
\begin{equation}
\eta_c = g_{\rm IB}^c/g_{\rm BB} \approx \sqrt{2} \pi c/g_{\rm BB} = \pi \sqrt{n_0 |a_{\rm BB}|}.
\label{eq:EtaCFrohlich}
\end{equation}
As indicated in Fig.~\ref{fig:Intro2}, the border of the weak coupling region scales as $\eta_c \sim \sqrt{n_0 |a_{\rm BB}|}$.

For the experimental parameters of Ref.~\cite{catani2012quantum}, the critical coupling strength where the Fr\"ohlich model breaks down is given by $\eta_c \approx 6$. Indeed, from Fig.~\ref{fig:1DpolaronMass} we see that theoretical calculations based on the Fr\"ohlich Hamiltonian only describe the experimental data for $\eta \lesssim 3$, and the deviations become large around $\eta_c$.

Condition \eqref{eq:EtaCFrohlich} provides an estimate when the Fr\"ohlich Hamiltonian is valid. Note that this is different from condition \eqref{eq:validityPerturbationTheory} which describes in which regime the lowest-order perturbation theory in boson-boson and impurity-boson interactions is valid. Comparison of the two expressions shows that there exists a parameter regime at large $n_0 |a_{\rm BB}|$ and $\eta < \eta_c$ where the Fr\"ohlich model is valid but has to be solved non-perturbatively.

\subsection{Feynman variational approach to the Fr\"ohlich model}
\label{subsec:FeynmanFrohlich}
In Refs.~\cite{Vlietinck2015,Grusdt2015RG,Shchadilova2016} the validity of Feynman's variational path integral description of Fr\"ohlich polarons has been questioned. When the impurity mass is small and interactions are moderate, a new regime has been identified where the correlations between phonons in the polaron cloud become important. Because Feynman's method merely interpolates between the two extremes of weak and strong coupling~\cite{Feranchuk2005}, it cannot capture the physics accurately in this situation.

\begin{figure}[t!]
\centering
\epsfig{file=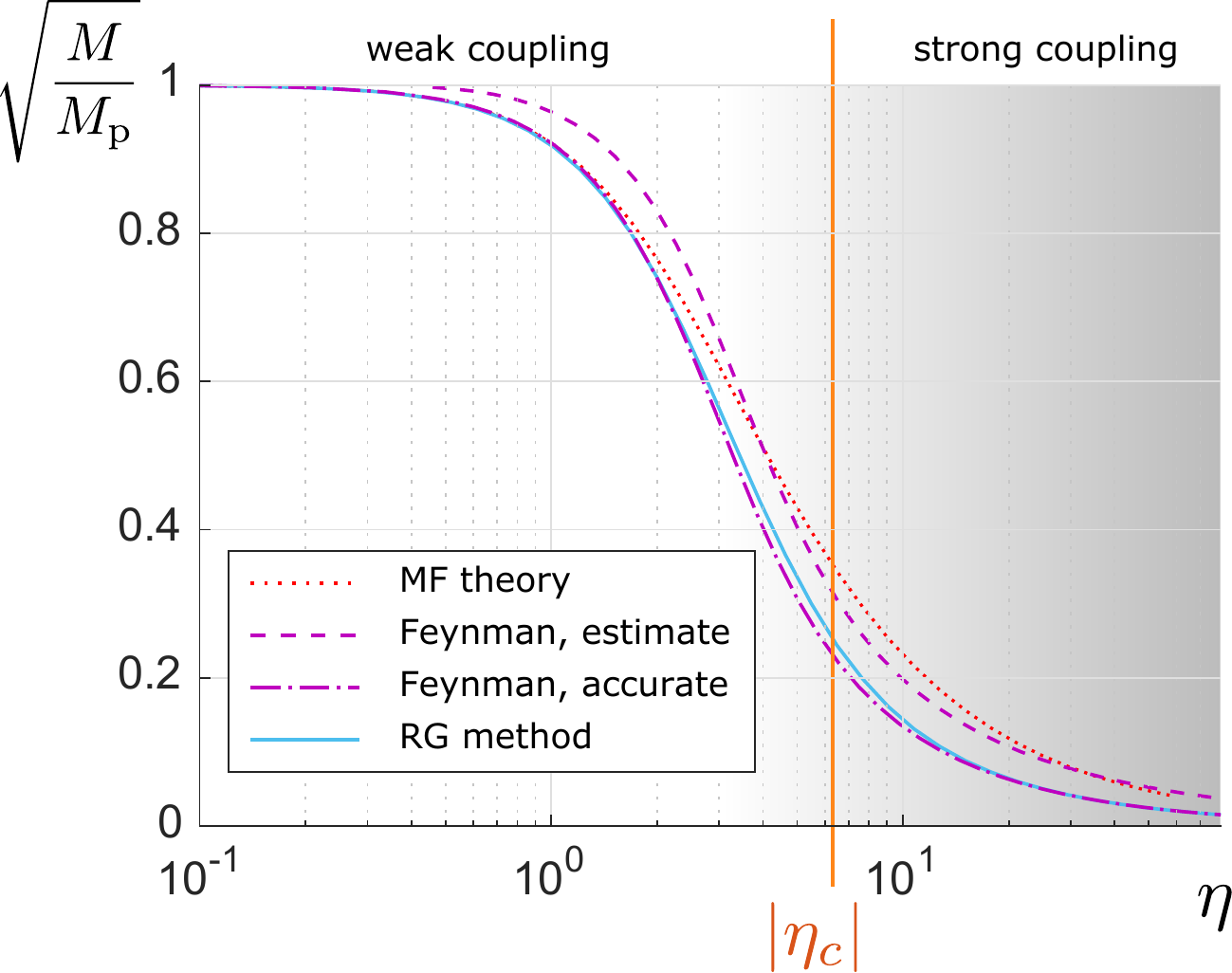, width=0.45\textwidth} $\quad$
\caption{Theoretical results for the effective polaron mass: We compare predictions by different theoretical approaches, all obtained assuming the Fr\"ohlich Hamiltonian $\H_F$. This model is only valid for $|\eta| \ll \eta_{\rm c}$ indicated in the plot. Parameters were chosen as in the experiment by Catani et al.~\cite{catani2012quantum}.}
\label{fig:polaronMassCompFroehlich}
\end{figure}

In the Florence experiment~\cite{catani2012quantum} the impurity mass is a factor of two smaller than the boson mass; It is then natural to ask how accurate Feynman's approach works in this case. To answer this question we compare predictions by the RG introduced in Ref.~\cite{Grusdt2015RG} to Feynman's variational path integral method, see Fig.~\ref{fig:polaronMassCompFroehlich}. Note that we only consider the Fr\"ohlich Hamiltonians $\H_F$ and $\H_F'$ in this figure. For the sake of comparison we also present results at strong couplings, beyond the critical value $\eta_c$ where the Fr\"ohlich model is not sufficient anymore.

To extract the effective polaron mass $M_{\rm p}$, Feynman suggested a simple estimate where one of his variational parameters serves as a direct approximation for the mass renormalization $\delta M = M_{\rm p} -M$. In addition he derived a more accurate expression for the polaronic mass enhancement from the imaginary-time Green's function, see Refs.~\cite{Feynman1955,Tempere2009,Casteels2012,Grusdt2015Varenna} for details. As can be seen from Fig.~\ref{fig:polaronMassCompFroehlich}, the estimated polaron mass has a large relative error, in particular in the regime of weak interactions.

The accurate expression for the polaron mass determined from the imaginary-time Green's function agrees remarkably well with our RG calculation. Only for intermediate couplings, small differences between the two methods can be observed. For weak and strong couplings both predictions coincide exactly. We conclude that Feynman's ansatz provides an accurate description of Fr\"ohlich polarons in the experiment by Catani et al.~\cite{catani2012quantum} when the Fr\"ohlich Hamiltonian $\H_F$ is used.

\section{Strong coupling: Effects of two-phonon terms}
\label{sec:TwoPhononTerms}
When the coupling $\eta \approx \eta_c$ becomes too large, the Fr\"ohlich model is no longer valid. In the following we will analyze the effect of two-phonon scattering terms $\H_{\rm 2ph}$ which modify the properties of the polaron in this regime~\cite{Rath2013}. We consider the beyond-Fr\"ohlich Hamiltonian defined by
\begin{equation}
\H_{\rm BF} = \H_F + \H_{\rm 2ph}.
\label{eq:defBeyFr}
\end{equation}
We compare our theoretical predictions based on $\H_{\rm BF}$ with results for the full microscopic model $\H$ obtained by Monte Carlo simulation.
\subsection{MF theory}
\label{subsec:MFtwoPhonon}
We start by highlighting some of the polaron properties specific to 1D systems, using MF formalism discussed in Ref.~\cite{Shchadilova2016PRL}. The formalism for 1D Bose polarons is presented in Appendix~\ref{sec:AppdxMF}. In short, firstly one utilizes conservation of the total momentum $p$ by applying the unitary transformation $\hat{U}_{\rm LLP}$ introduced by Lee, Low and Pines~\cite{Lee1953}. This transforms the original Hamiltonian $\H_{\rm BF}$ into the new one, $\tilde{\mathcal{H}}_{\rm BF} = \hat{U}_{\rm LLP} \H_{\rm BF} \hat{U}_{\rm LLP}^\dagger$. Next, a product wave function of coherent phonon states is assumed to describe the ground state of $\tilde{\mathcal{H}}_{\rm BF}$. This means that the polaron state can be written as 
\begin{equation}
\ket{\psi_{\rm MF}} = \hat{U}_{\rm LLP}^\dagger \prod_{k} \ket{\alpha_{k}^{\rm MF}}.
\end{equation}

To find the minimum variational energy $E^{\rm MF}[\alpha_k] = \bra{\psi_{\rm MF}} \H_{\rm BF} \ket{\psi_{\rm MF}}$ one solves the saddle point equations $\delta E_{\rm MF}[\alpha_k] / \delta \alpha_k = 0$, which yields
\begin{equation}
\alpha_{k}^{\rm MF} = - \frac{\beta_\MF V_k}{\omega_k + k^2/2M - k ( p - P_{\rm ph}^\MF ) / M }.
\end{equation}
The total phonon momentum $P_{\rm ph}^\MF = \int dk ~ k |\alpha_{k}^{\rm MF}|^2$ and $\beta_\MF$ have to be determined self-consistently. The MF variational polaron energy is given by
\begin{equation}
E_0^\MF = n_0 g_\IB \beta_\MF + \frac{g_\IB}{2 \pi} \int dk ~ \sinh^2 \theta_k  + \frac{p^2 - (P_\ph^\MF )^2}{2M}.
\label{eq:EoMF}
\end{equation}
Note that MF calculations go beyond a straightforward perturbative treatment of impurity-boson interactions as presented e.g. in Ref.~\cite{Parisi2016}.

\begin{figure}[b!]
\centering
\epsfig{file=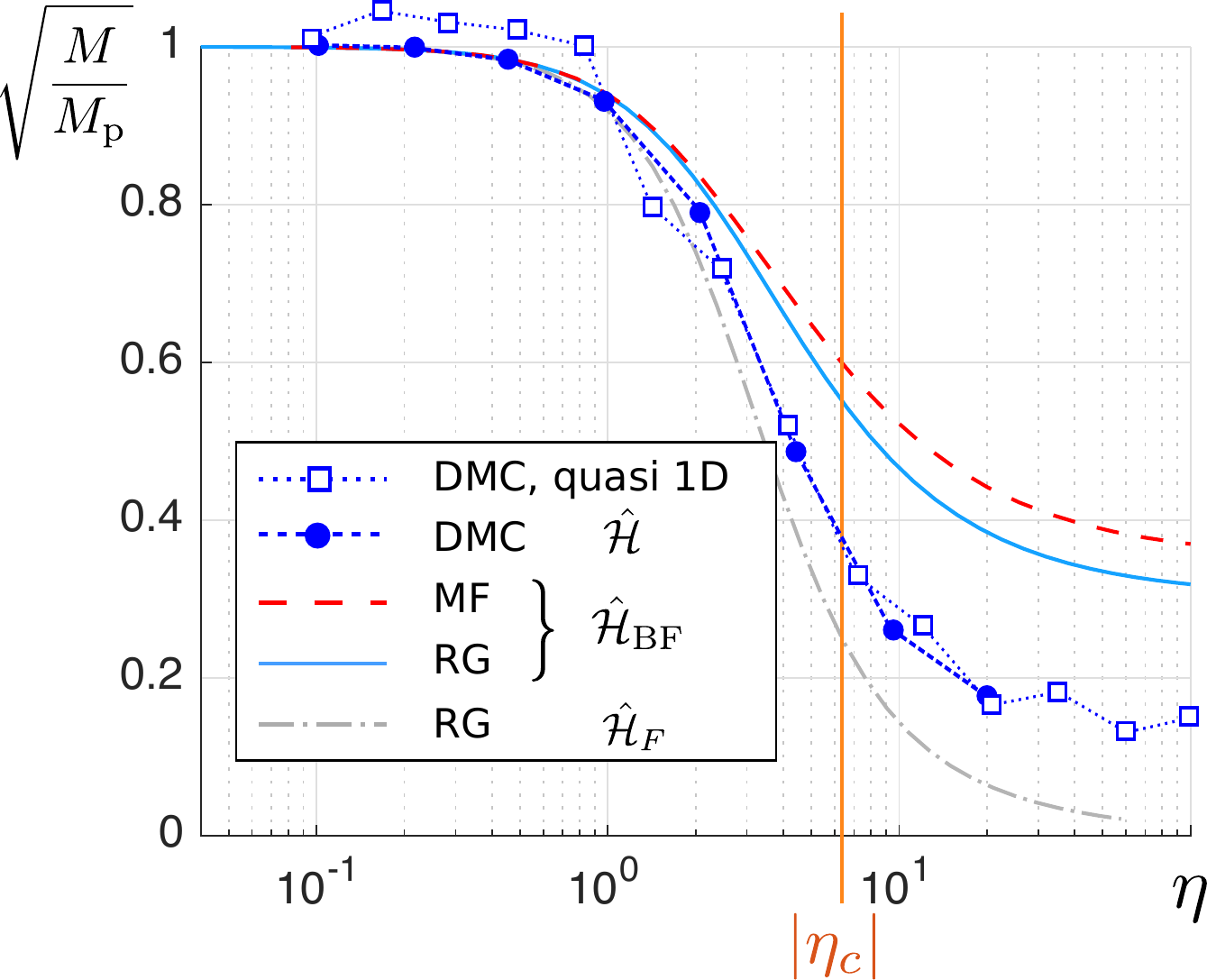, width=0.4\textwidth} $\qquad$
\caption{The effective polaron mass $M_{\rm p}$ is compared for different model Hamiltonians and using different theoretical methods, as indicated in the legend. The Fr\"ohlich model is only valid for $|\eta| \ll |\eta_c|$. Parameters are chosen as in the experiment by Catani et al. \cite{catani2012quantum}, and all DMC calculations were performed for $N=50$ particles. The quasi-1D calculations are computationally more demanding, explaining the increased amount of noise in the data.}
\label{fig:polaronMassComp2ph}
\end{figure}

\emph{Effective polaron mass.--}
The effective mass of the polaron, $M_{\rm p}$, can be obtained from the momentum dependence of the MF polaron energy. As discussed e.g. in Ref.~\cite{Grusdt2015Varenna}, one finds
\begin{equation}
E_0^\MF(p) = \frac{p^2}{2 M_{\rm p}} + \mathcal{O}(p^4).
\end{equation}
In Fig.~\ref{fig:1DpolaronMass} we compare the MF effective polaron mass $M_{\rm p}$ (dashed line) to the experimental results. At large repulsive couplings around $\eta \approx \eta_c$ we find a saturation of $M_{\rm p}$ at some large but finite value. Qualitatively, a similar saturation of the polaron mass has been observed in the measurements by Catani et al.~\cite{catani2012quantum}, although it remains unclear how reliable the utilized measurement procedure is at strong couplings, see Sec.~\ref{subsec:polaronOsc2} for a discussion. 

Our most reliable DMC simulations in finite-size systems also show a saturation of the effective polaron mass at strong interactions $g_{\rm IB} \gg g_{\rm BB}$, see Fig.~\ref{fig:1DpolaronMass}. The saturated value $M_{\rm p}(\eta \to \infty)$ is substantially smaller than the result expected from MF theory. In addition, from finite-size scaling we find that the DMC results are consistent with an infinite polaron mass in the thermodynamic limit. Indeed, based on the microscopic model Hamiltonian in Eq.~\eqref{eq:Hmicro} we expect that $M_{\rm p} \propto N$ is proportional to the number of bosons $N$ when impurity-boson interactions diverge, $g_{\rm IB} \to \infty$. To understand this, note that the average impurity and boson velocities $v_{\rm imp} = v_{\rm B}$ are equal because impurities and bosons cannot penetrate each other. By relating them to the average impurity and boson momenta $p_{\rm I} = M v_{\rm I}$ and $p_{\rm B} = m_{\rm B} v_{\rm B}$ using Ehrenfest's theorem we see that a finite total momentum $p = p_{\rm I} + p_{\rm B} = \mathcal{O}(N^0)$ is distributed over all particles and it follows that $v_{\rm I} = \mathcal{O}(1/N)$. Because the average impurity velocity $v_{\rm I} = v_{\rm p}$ coincides with the average polaron velocity $v_{\rm p} = p / M_{\rm p}$, it follows that $M_{\rm p} = \mathcal{O}(N)$. We also note that an explicit proof that $M_{\rm p} \propto N$ has been given by means of exact Bethe-ansatz calculations in the limit when both impurity-boson and boson-boson interactions diverge, $g_{\rm IB} = g_{\rm BB} \to \infty$, see Ref.~\cite{Fuchs2005}.

The MF theory is based on the Bogoliubov approximation and ignores phonon-phonon interactions. Thus it cannot fully capture the impenetrable nature of impurities and bosons in the limit when $g_{\rm IB} = \infty$ in Eq.~\eqref{eq:Hmicro}. As discussed in Appendix~\ref{appdx:RGapproach} this leads to an RG flow of the interaction strength which shifts the position of the resonance defined by the divergence of impurity-boson interactions at long wavelengths. MF theory predicts an infinite effective polaron mass at an attractive microscopic interaction strength $\eta_{\rm c, MF} < 0$, as indicated in Fig.~\ref{fig:1DpolaronMass}. When boson-boson interactions are fully taken into account, as in our DMC simulations, such shifts of the impurity-boson resonance cannot occur. 

Experimentally \cite{catani2012quantum}, the impurity-boson interaction $g_{\rm IB}$ can be tuned by utilizing a confinement-induced (Feshbach) resonance \cite{Olshanii1998}. When formulating the effective Hamiltonian in Eq.~\eqref{eq:Hmicro}, we implicitly assumed that the scattering of an impurity with a boson at low energies is not modified by many-body effects caused by the interactions with the surrounding Bose gas. Similar to the shift of the resonance position predicted by the MF and RG theories, we expect that many-body effects can lead to a shift of the confinement-induced resonance. 

In Fig.~\ref{fig:polaronMassComp2ph} we also show results from a quasi-1D DMC calculation. It takes into account the finite extent of the trap in radial direction. The prediction for the effective polaron mass is in excellent agreement with our strictly-1D DMC calculations. This suggests that the large discrepancy between theory and experiment in the strong coupling regime cannot be explained by the influence of radially excited states in the trap. Hence we conclude that the relevant physics remains strictly one-dimensional. Details of this calculation can be found in Sec.~\ref{subsec:DMCapproach}.

\emph{Multi-particle bound states and in-medium Feshbach resonance.--}
For a situation where the polaron momentum $p=0$ vanishes, $\beta_\MF$ is given by Eq.~\eqref{eq:betaPzero}. This mean-field expression suggests the existence of a transition where $\beta_{\rm MF} \to \infty$ diverges. It takes places at a critical attractive coupling strength
\begin{equation}
g_{\IB,c}^{\MF} = - 2 \pi \left[  \int dk ~ \frac{W_k^2}{\omega_k + k^2/2M}  \right]^{-1} < 0.
\label{eq:gIBcMF}
\end{equation}
At this interaction strength the number of phonons in the polaron diverges, see Eq.~\eqref{eq:NphMF}. This suggests an instability of the system towards a state with a large number of bosons accumulating around the impurity. Such behavior has been associated with multi-particle bound states in higher dimensions \cite{Ardila2015,Shchadilova2016PRL,Grusdt2016RGBEC}.

\begin{figure}[t!]
\centering
\epsfig{file=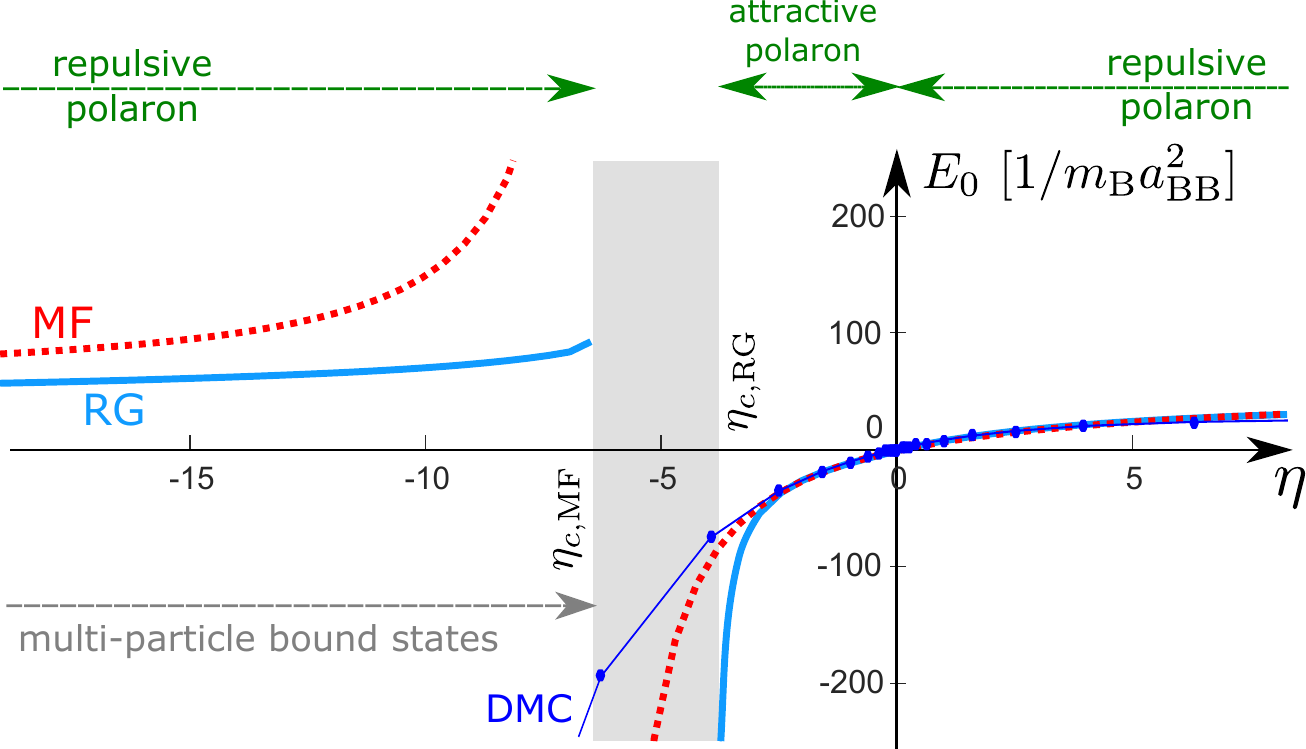, width=0.48\textwidth}
\caption{The polaron energy calculated from MF theory and using the RG method, for parameters as in the Florence experiment~\cite{catani2012quantum}. For comparison, results from our DMC calculations are shown for which we performed extrapolations to the thermodynamic limit $N \to \infty$, see Appendix \ref{appdx:finite-size} for details. Different regimes discussed in the text are indicated in the top row. In the shaded area the RG approach breaks down because the number of bosons in the polaron cloud diverges. Note that DMC calculations fully include all microscopic terms in the Hamiltonian, in particular boson-boson interactions which lead to a stable solution in the regime where the RG breaks down. Here the true ground state is a correlated state of many interacting bosons accumulating around the impurity. For weak attractive interactions, $\eta_{\rm c, RG} < \eta < 0$, an attractive polaron exists. On the other hand, for strong attractive interactions, $\eta<\eta_{\rm c, MF}$, two branches can be realized. The energetically lower branch (energies not calculated) contains multi-particle bound states \cite{McGuire1964,Ardila2015,Shchadilova2016PRL}, while the energetically higher branch corresponds to a repulsive polaron with attractive microscopic interactions. For more details on the phase diagram, see Appendix~\ref{appdx:RGapproach} and Ref.~\cite{Grusdt2016RGBEC}.}
\label{fig:1DphaseDiag}
\end{figure}

\begin{figure*}[htb]
\centering
\epsfig{file=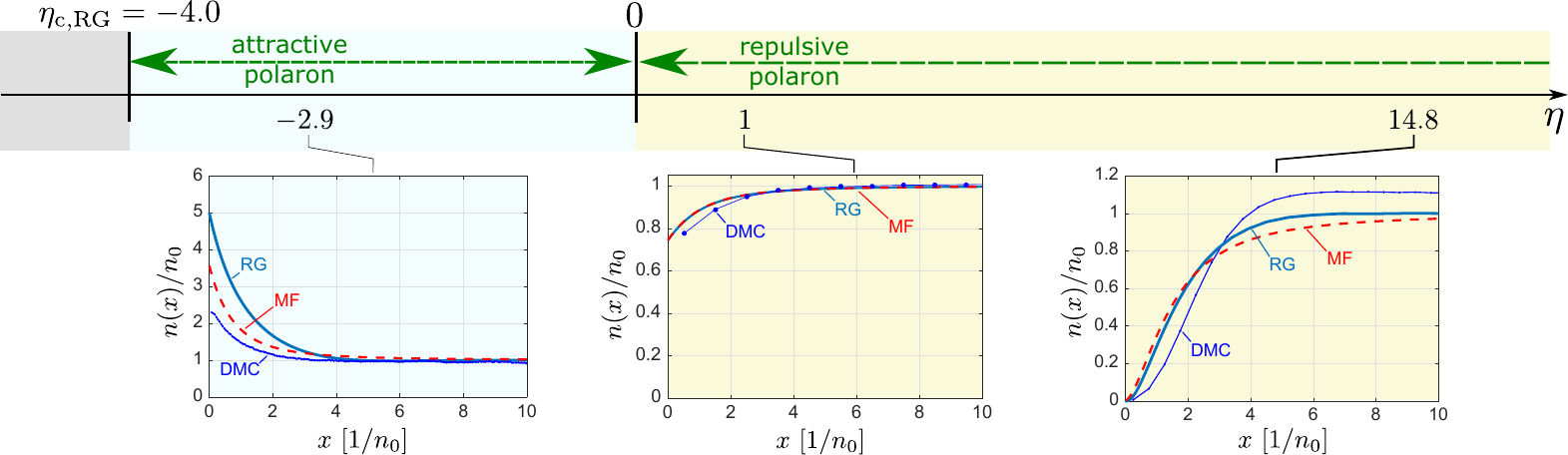, width=0.94\textwidth}
\caption{Density profile around the impurity: The boson density around the impurity (at $x=0$) is calculated using the MF (dashed) and RG (solid) methods and compared to DMC simulations (dots) for $N=50$ bosons. The parameters correspond to the experiment by Catani et al.~\cite{catani2012quantum}. RG and MF calculations are based on the beyond-Fr\"ohlich Hamiltonian $\H_{\rm BF}$, which neglects phonon-phonon interactions. Their effect leads to the observable discrepancies in comparison to DMC calculations which fully include the boson-boson interactions.}
\label{fig:depletion}
\end{figure*}

At $g_{\IB,c}^{\MF}$ the MF polaron energy also diverges and changes sign, see Eq.~\eqref{eq:EoMF}. This indicates a transition from an attractive polaron with negative energy, to a repulsive polaron with positive energy. In Fig.~\ref{fig:1DphaseDiag} the MF polaron energy is shown and the MF critical value $\eta_{c, \MF} = g_{\IB,c}^{\MF} /g_{\rm BB}$ is indicated in the plot. In addition, we calculate the density profile of the bosons in Fig.~\ref{fig:depletion}. When $\eta_{c, \MF}$ is approached from $\eta=0$, as expected, a large number of bosons accumulates around the impurity. This effect is also observed by our full numerical Monte Carlo calculations.

To understand the physics of this transition, we first analyze the limits $g_{\rm IB} \to \pm \infty$. Because the MF wavefunction coincides for $g_{\rm IB}= + \infty$ and $g_{\rm IB}= - \infty$, we conclude that the two repulsive polaron branches at strong attraction and repulsion are adiabatically connected. This behavior is reminiscent of the meta-stable super-Tonks-Girardeau state, which can be realized by a strongly interacting Bose gas (without impurities) when the interactions are quickly changed from strongly repulsive to strongly attractive \cite{Astrakharchik2005,Haller2009}.

For an impurity inside a non-interacting Bose gas, $g_{\rm BB}=0$, the critical value becomes zero, $g_{\IB,c}^{\MF} = 0$. In this case, $g_{\IB,c}^{\MF}$ corresponds to the point where we expect the appearance of an infinite series of multi-particle impurity-boson bound states in 1D, if the impurity mass is infinite. This last condition guarantees that no correlations can be induced between the bosons by scattering off the impurity. Note that such processes are not included in the MF wavefunction, even for finite impurity mass $M < \infty$. This explains why $g_{\IB,c}^{\MF} = 0$ vanishes in the non-interacting Bose gas, independent of $M$. In contrast, the RG approach presented in Appendix \ref{appdx:RGapproach} includes impurity-induced interactions between the phonons. There we show that $g_{\IB,c}^{\rm RG} \neq 0$ becomes non-zero in the non-interacting Bose gas when the impurity mass is finite.

In the limit of a non-interacting Bose gas, $g_{\rm BB}=0$,  and for a localized impurity, $M=\infty$, we note that the 1D scattering length $a_{\IB} = - 1 / m_{\rm B} g_\IB$~\cite{Bloch2008} diverges when $g_\IB$ approaches the critical value $g_{\IB,c}^{\MF} = 0$. This effect can be associated with a 1D Feshbach resonance. On the other hand, when the Bose gas is interacting or the impurity becomes mobile, the position of the Feshbach resonance is shifted. It is now located at the bare two-particle interaction $g_{\IB,c}$. This behavior is reminiscent of the in-medium shift of the Feshbach resonance predicted for Bose polarons in 3D. In this case, too, the Feshbach resonance is associated with the appearance of multi-particle bound states in the spectrum~\cite{Shchadilova2016PRL,Grusdt2016RGBEC}.

In general one should be cautious that the inclusion of quantum fluctuations might change the position of the transition compared to the MF prediction. In some cases, transitions predicted by MF theory even disappear completely. We address this problem in Appendix~\ref{appdx:RGapproach}, where quantum fluctuations are included using an RG approach. See also Ref.~\cite{Grusdt2016RGBEC} for a detailed discussion.

\emph{Logarithmic IR divergence of the polaron energy.--}
In contrast to the 3D case~\cite{Shchadilova2016PRL}, expression~\eqref{eq:EoMF} for the MF polaron energy is fully convergent when the large-momentum cut-off $\Lambda_0$ is sent to infinity. It has a logarithmic divergence when the small-momentum cut-off $\lambda$ is sent to zero however,
\begin{equation}
\frac{g_{\rm IB}}{2 \pi} \int_\lambda^{\Lambda_0} dk ~ \sinh^2 \theta_k \sim - g_\IB \log \lambda  \stackrel{\lambda \to 0}{\longrightarrow} \rm{sgn} (g_\IB) \times \infty.
\label{eq:e0MFlogDiv}
\end{equation}
This IR divergence can be regularized by including quantum fluctuations using the RG approach presented in Sec.~\ref{subsec:RGapproach}. 

In our MF calculations of the polaron energy, for example in Fig.~\ref{fig:1DphaseDiag}, we ignore the log-divergent term $g_{\rm IB} / 2 \pi \int dk~ \sinh^2 \theta_k$ in Eq.~\eqref{eq:EoMF}. This step can only be justified a-posteriori, by showing that the general behavior of the MF polaron energy without the log-divergent term closely resembles the fully regularized RG prediction. Alternatively one can argue that the divergence $\sim \log \lambda$ is only logarithmic in system size $L \sim 1/ \lambda$. Hence the correction to the polaron energy from the log-divergent term in Eq.~\eqref{eq:e0MFlogDiv} is not expected to be large for a finite system of an experimentally relevant size. As a consequence, our MF prediction for the polaron energy is not expected to be quantitatively accurate.

\emph{Orthogonality catastrophe.--}
In one dimension, the MF theory of Bose polarons has even more noteworthy peculiarities associated with the IR cut-off. First of all, the phonon number in the polaron cloud diverges logarithmically,
\begin{equation}
N_\ph^\MF= \int_\lambda dk ~ \l \frac{ \beta_\MF V_k}{\Omega_k^\MF} \r^2  \sim - \log \lambda \stackrel{\lambda \to 0}{\longrightarrow}  \infty.
\end{equation}
This is directly related to the log-divergence of the MF energy. Depleting the quasi-1D condensate and creating infinitely many phonons costs an interaction energy which scales like $\sim g_{\rm IB} N_{\rm ph}$.

The diverging phonon number can also be understood as a manifestation of Anderson's orthogonality catastrophe~\cite{Anderson1967} for a mobile impurity in a 1D Bose gas. Within MF theory, the quasiparticle residue is determined
by the simple relation (see e.g. Ref.~\cite{Grusdt2015Varenna})
\begin{equation}
Z=e^{- N_{\rm ph}^\MF} \simeq \lambda \to 0.
\end{equation}
Therefore already an infinitesimally small interaction leads to a vanishing quasiparticle weight $Z=0$ in an infinite system.

\emph{Supersonic polarons.--}
The large number of phonons in the 1D polaron cloud also affects the dependence of polaron properties on the total momentum \cite{Schecter2012}. While there exists a phase transition from the subsonic to the supersonic regime at large momenta in higher dimensions~\cite{Shashi2014RF}, this transition is absent in 1D. To calculate the critical momentum $p_c^\MF$ where the subsonic to supersonic transition for the impurity takes place, we employ Landau's criterion for superfluidity. It states that the transition takes place when the polaron velocity $v_{\rm p}$ exceeds the speed of sound $c$ in the Bose gas. From the expression $v_{\rm p} = (p - P_\ph^\MF) / M$ \cite{Shashi2014RF} we derive
\begin{equation}
p_c^\MF = M c + \int dk \frac{k \beta_\MF^2 V_k^2}{\l \omega_k + k^2/2 M - k c \r^2} \simeq \frac{1}{\lambda} \to \infty.
\end{equation}
Here $\lambda$ is the IR momentum cut-off. We find that polarons are subsonic in 1D, due to the dressing of the impurity with an infinite number of phonons.

\subsection{RG approach}
\label{subsec:RGapproach}
In Appendix~\ref{appdx:RGapproach} we extend our analysis and include quantum fluctuations on top of the MF solution by applying the non-perturbative RG approach introduced in Refs.~\cite{Grusdt2015RG,Grusdt2016RG,Grusdt2016RGBEC}. The RG procedure allows us to regularize the IR divergence of the polaron energy identified in Eq.~\eqref{eq:e0MFlogDiv}. We use this method to make predictions for the energy shift between an interacting and a free impurity, which can be {measured directly using radio-frequency spectroscopy \cite{Jorgensen2016PRL,Hu2016PRL}. On the other hand, the RG method provides new insights into the polaron phase diagram. We present a detailed discussion of our RG analysis of 1D Bose polarons in Appendix~\ref{appdx:RGapproach}.

From the solution of the RG flow equations (see Appendix~\ref{appdx:RGapproach}) we obtain the fully regularized polaron energy $E_0^{\rm RG}$. Explicit calculation in the appendix demonstrates IR divergence of the polaron energy is regularized with this approach. We further benchmark the RG approach by comparing to the DMC calculations and working in the regime of large boson density $n_0$. More concretely, we assume that the dimensionless interaction strength $\gamma \ll 1$, see Eq.~\eqref{eq:defGamma}, is small. In this regime the Bogoliubov approximation is justified and the effect of phonon-phonon interactions, included only by DMC, is expected to be weak.

We estimate the importance of non-linear interaction terms between Bogoliubov phonons by the relative size of corrections $\Delta \varepsilon_{\rm LHY}$ to the ground state energy $\varepsilon_0$ of the homogeneous Bose gas caused by quantum fluctuations \cite{Lieb1963a,Lieb1963}. Similar to Lee-Huang-Yang corrections in 3D~\cite{Lee1957,Lee1957a}, one obtains
\begin{equation}
\Delta \varepsilon_{\rm LHY} / \varepsilon_0 = - \frac{2^{5/2}}{3 \pi} \frac{1}{\sqrt{n_0 |a_{\rm BB}|}} = -\frac{4}{3 \pi} \sqrt{\gamma},
\label{eq:defLHY}
\end{equation}
see e.g. Ref.~\cite{Bloch2008}. In the second expression, $\gamma$ is the dimensionless interaction strength in the Bose gas, see Eq.~\eqref{eq:defGamma}.

\begin{figure}[t!]
\centering
\epsfig{file=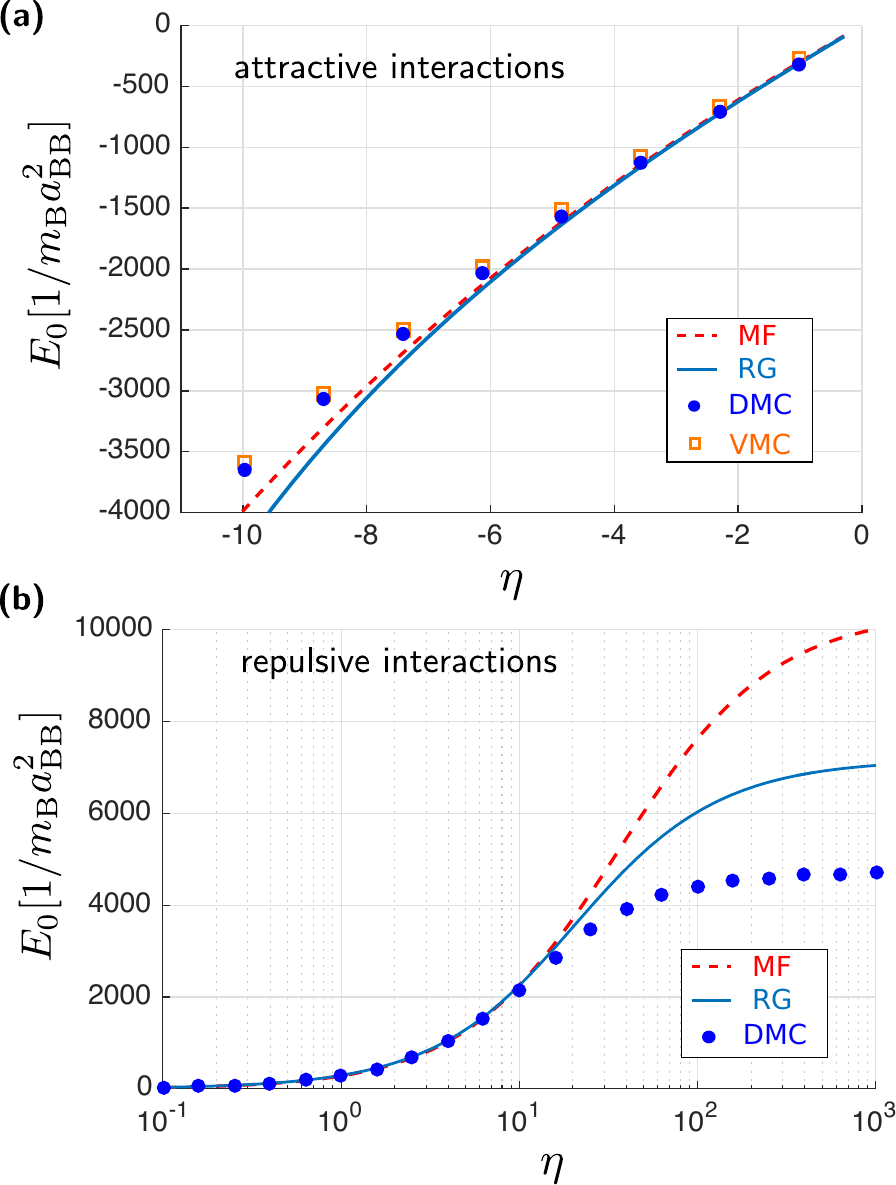, width=0.44\textwidth} $\quad$
\caption{We compare the polaron energy, calculated from different methods, for attractive (a) and repulsive (b) microscopic interactions. We used a value of $\gamma=0.014$ (or $n_0|a_{\rm BB}| = 144$) corresponding to the weakly interacting regime, where Bogoliubov theory is expected to be applicable. Parameters are chosen similarly to those of Florence experiment~\cite{catani2012quantum}, except that we are assuming a larger boson density. The DMC results are obtained by $1/N$ extrapolation to the thermodynamic limit $N \to \infty$, see Appendix \ref{appdx:finite-size} for details.}
\label{fig:compHighDensityRep}
\end{figure}

In Fig.~\ref{fig:compHighDensityRep} we compare our result for the polaron energy to DMC calculations. We used the same mass ratio $M/m_{\rm B} = 41/87$ as in the experiment by Catani et al.~\cite{catani2012quantum}, but the density was chosen to be $n_0 |a_{\rm BB}|=144$ (corresponding to $n_0 =6/\xi=144/|a_{\rm BB}|$ or $\gamma=0.014$) so that the corrections of quantum fluctuation to the energy is $\Delta \varepsilon_{\rm LHY} / \varepsilon_0 = 0.05 \ll 1$. For weak-to-intermediate couplings we obtain excellent agreement between RG and DMC predictions, validating the use of the effective beyond-Fr\"ohlich Hamiltonian $\H_{\rm BF}$ in this regime. For the case of repulsive microscopic interactions, $g_{\rm IB}>0$, we obtain sizable deviations in the regime of very strong couplings. Yet the correction by the RG accounts for about half of the deviation between MF and DMC results in this case.

In the case of attractive microscopic interactions $g_{\rm IB}<0$, see Fig.~\ref{fig:compHighDensityRep} (a), an additional regularization of the polaron energy $E_0(\Lambda)$ is required in the RG to deal with divergencies of one of the coupling constants ($G_-$) during the RG. This procedure is described in detail in Ref.~\cite{Grusdt2016RGBEC}. The agreement for the energy of attractive polarons is reasonable. It should also be noted that the attractive polaron is not expected to be the ground state of the Hamiltonian $\H_{\rm BF}$ based on the Bogoliubov approximation for the Bose gas in this case, as discussed in detail in Appendix~\ref{appdx:RGapproach}. Instead it has been argued in Ref.~\cite{Grusdt2016RGBEC} that the true ground state contains a large number of bosons accumulating around the impurity. This prediction is consistent with our DMC calculations in this regime.

In Fig.~\ref{fig:Intro} (b) we calculate how the density profile of the bosons is modified by impurity-boson interactions.
Again we consider the high-density regime where the Bogoliubov approximation for the Bose gas is justified. In contrast to the energies shown in Fig.~\ref{fig:compHighDensityRep}, the depletion of the Bose gas is accurately described by the RG approach in this case, even for very large interactions ($\eta=10^3$ in the figure). This serves as another important benchmark for the RG and the use of the effective beyond-Fr\"ohlich Hamiltonian $\H_{\rm BF}$ in the regime where quantum fluctuations in the bath are weak.

In contrast, for the experiment by Catani et al.~\cite{catani2012quantum} the correction to the energy due to quantum fluctuations is sizable, $\Delta \varepsilon_{\rm LHY} / \varepsilon_0 \approx -0.30$, indicating that Bogoliubov theory is no longer quantitatively accurate. Indeed the predictions for the effective polaron mass in the strong coupling regime differ by a factor of $4$, see Fig.~\ref{fig:polaronMassComp2ph}. Below, in Fig.~\ref{fig:polaronEnergyStrongCoupling}, we also compare predictions for the polaron energy in this case and find large quantitative deviations between the two approaches on the repulsive side.

\section{Dynamics of strong coupling polarons}
\label{sec:PolaronDynamics}
Comparison of the effective polaron mass in Figs.~\ref{fig:1DpolaronMass} and~\ref{fig:polaronMassComp2ph} suggests a good agreement between theory and experiment for weak and intermediate interaction strength. However, in the strong coupling regime the value of the plateau is not reproduced. Here we investigate the measurement method itself, which relies on the analysis of oscillations of impurities subject to a shallow trapping potential. We show that inside a homogeneous gas, both their amplitude and frequency renormalization can be used to measure the effective mass. Surprisingly, no frequency renormalization was observed by Catani et al.~\cite{catani2012quantum}, who investigated the reduction of the amplitude instead. We speculate here that this can be caused by the inhomogeneity of the Bose gas. Indeed, its size is below the extent of the largest oscillation amplitudes contributing to the measurement of the effective mass.

\subsection{Time-dependent MF theory}
\label{subsec:tDepMF}
To describe the dynamics of strongly coupled Bose polarons inside a shallow trapping potential, we employ a time-dependent variant of MF theory as in Refs.~\cite{Shashi2014RF,Shchadilova2016PRL}. We supplement this approach with the local-density approximation (LDA). This allows us to treat the problem in the Lee-Low-Pines frame despite the external potential.

Our variational ansatz corresponds to a product wave function of coherent states,
\begin{equation}
\ket{\psi_\MF(t)} = e^{- i \chi(t)} \hat{U}_{\rm LLP}(P(t))  \prod_{k} \ket{\alpha_{k}(t)},
\end{equation}
in the frame which is co-moving with the impurity, see Appendix~\ref{sec:AppdxMF}. This is achieved by the unitary transformation $\hat{U}_{\rm LLP}(P(t))$ \cite{Lee1953}, where $P(t)$ denotes the time-dependent total system momentum. $\chi(t)$ denotes an overall phase which guarantees conservation of the total energy. Similar to the MF ground state, this wavefunction  ignores quantum fluctuations of phonons induced by the mobile impurity.

The position of the impurity $X(t) = \langle \hat{x} \rangle$ can be calculated in the LLP frame by using the Ehrenfest theorem,
\begin{equation}
\frac{d}{dt} X(t) = \l P(t) - P_{\rm ph}(t) \r / M,
\end{equation}
where $P_{\rm ph}(t) = \int dk~ k |\alpha_k(t)|^2$ is the expectation value of the phonon momentum at time $t$. The equations of motion for $\alpha_k(t)$ can be derived from the Hamiltonian $\H_{\rm BF}$, see Eq.~\eqref{eq:defBeyFr}, using Dirac's time-dependent variational principle~\cite{Jackiw1980}. As discussed in Ref.~\cite{Shchadilova2016PRL} this yields
\begin{multline}
i \partial_t \alpha_k = V_k + \biggl( \omega_k + \frac{k^2}{2 M} - \frac{k}{M} (P(t) - P_{\rm ph}(t)) \biggr) \alpha_k \\
+ \frac{g_\IB}{2 \pi} \int dk'  \left[ {\rm Re} (\alpha_{k'}) W_k W_{k'}  + i ~ \frac{{\rm Im} (\alpha_{k'})}{W_k  W_{k'} } \right].
\label{eq:EOMalphaMFt}
\end{multline}

We note that the log-divergent term in the MF polaron energy, see Eq.~\eqref{eq:e0MFlogDiv}, does not enter in the equations of motion~\eqref{eq:EOMalphaMFt}. Therefore the time-dependent MF theory is well-behaved in the long-wavelength limit. We note that starting from a free impurity we observe numerically that the phonon number grows logarithmically in time,
\begin{equation}
N_{\rm ph}(t) \simeq \log t.
\end{equation}
This can be understood as a dynamical manifestation of the orthogonality catastrophe in a 1D Bose gas. In the polaron ground state the number of phonons diverges logarithmically even after inclusion of quantum fluctuations by the RG, see Appendix \ref{subsubsec:IRlogDivRegRG}. By analogy we expect that the dynamical divergence of the phonon number in time is not an artifact of the MF theory but a physical effect observable also in the presence of quantum fluctuations.

\subsection{Local-density approximation}
\label{subsec:LDA}
In the absence of an external force, the total system momentum is conserved, $d P(t) /dt=0$. Now we consider the effect of an additional (confinement) potential $V(x)$ on the impurity. In the limit when its changes are small over a length scale set by the size of the polaron wavepacket, we can still treat the system as a homogeneous one. In this LDA, the equation of motion for the total momentum is given by
\begin{equation}
\frac{d}{dt} P(t) = - \partial_x V(X(t)).
\end{equation}

\emph{Potentials from the Florence experiment.--}
In the following we will consider different potentials, which have all played a role in the experiment by Catani et al.~\cite{catani2012quantum}. Firstly, both the impurity and the bosons are trapped inside a shallow parabolic potential,
\begin{equation}
V_{\rm I, B}(x) = \frac{1}{2} m_{\rm I,B} \Omega_{\rm I,B}^2 x^2,
\end{equation}
where $m_{\rm I} = M$ is the impurity mass and $\Omega_{\rm I,B}$ are the trapping frequencies. The oscillator lengths in Ref.~\cite{catani2012quantum} are $\ell_{\rm I} \approx 11 \xi$ and $\ell_{\rm B} \approx 9 \xi$ respectively, justifying the LDA in this case because the size of the polaron is of the order $\xi \ll \ell_{\rm I,B}$.

In addition, Catani et al.~\cite{catani2012quantum} used a tight species-selective dipole trap (SSDP) to localize the impurities in the beginning. This gives rise to a steeper harmonic trap, $V_{\rm SSDP}(x)=M \omega_{\rm I}^2 x^2/2$, where $\omega_{\rm I} = 0.28 c/\xi$ in our units. In this case the oscillator length is $\ell \approx 3.3 \xi$ and the LDA is less justified.

Finally, the impurity energy depends on the density $n(x)$ of the Bose gas.
This gives rise to an effective potential which is determined by the interaction strength $g_\IB$. Assuming that the impurity can adiabatically follow its ground state at $P=0$, we obtain the following polaronic potential,
\begin{equation}
V_{\rm pol}(x) \equiv E_0^{\rm MF}(x) = g_\IB \beta_\MF(x) n(x).
\label{eq:Vpolx}
\end{equation}
To calculate $\beta_\MF(x)$ we note that the healing length $\xi$ and the speed of sound $c$ both depend on the density $n(x)$, and thus on $x$. For a weakly interacting quasi-1D condensate in a shallow trap we can apply the Thomas-Fermi approximation to estimate the boson density,
\begin{equation}
n(x) \approx n_0 - V_{\rm B}(x) / g_{\rm BB}.
\end{equation}

\begin{figure}[t!]
\centering
\epsfig{file=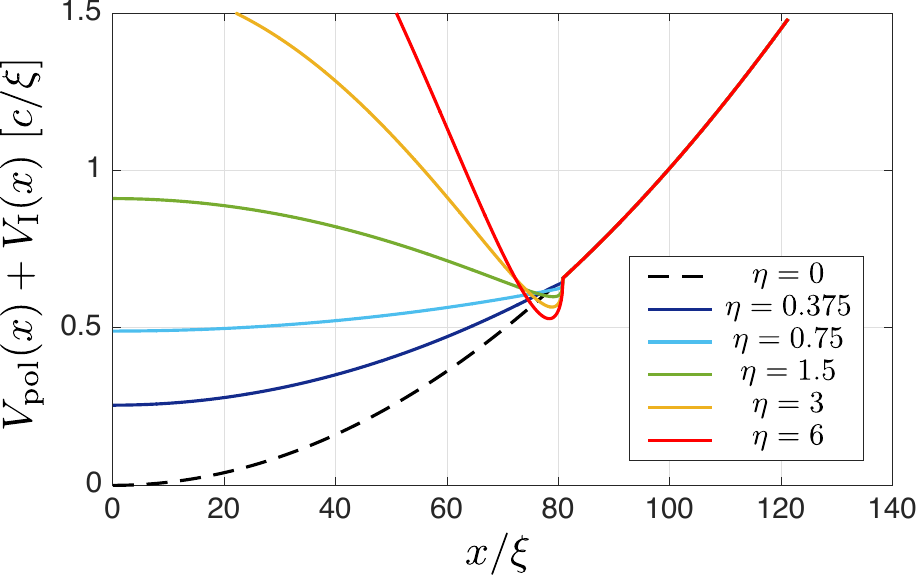, width=0.47\textwidth}
\caption{The effective potential, including $V_{\rm I}(x)$ (dashed line) and the polaronic part $V_{\rm pol}(x)$, is calculated for parameters as in the experiment by Catani et al.~\cite{catani2012quantum}. The size of the Bose gas was on the order of $R \approx 120 \xi$ in Ref.~\cite{catani2012quantum}.}
\label{fig:EffPotential}
\end{figure}

In Fig.~\ref{fig:EffPotential} the potentials are shown for parameters as in the experiment by Catani et al.~\cite{catani2012quantum} and assuming repulsive impurity-boson interactions. We note that in the strong coupling regime, for $\eta \gtrsim 1$, the effective polaronic potential $V_{\rm pol}(x)$ has its minimum at the edge of the Bose gas. In this case we expect that impurities become trapped in this region after they have had enough time to equilibrate with the quasi-1D condensate. Such behavior has indeed been observed in Ref.~\cite{catani2012quantum} for $\eta \gtrsim 1$.

\subsection{Polaron oscillations: homogeneous Bose gas}
\label{subsec:polaronOsc1}
We begin by studying the dynamics of an impurity interacting with a homogeneous Bose gas, where the impurity is subject to a shallow trapping potential. Experimentally this situation corresponds to the assumption that $\Omega_{\rm B} \ll \Omega_{\rm I}$ is small. In this case we can ignore the polaronic potential in Eq.~\eqref{eq:Vpolx}, and we will consider only $V_{\rm I}(x)$ now.

\emph{Adiabatic limit.--}
In the limit where the polaron follows its local ground state adiabatically, it can be described by an effective Hamiltonian
\begin{equation}
\H_{\rm eff} = \frac{\hat{p}^2}{2 M_{\rm p}} + \frac{1}{2} M \Omega_{\rm I}^2 \hat{x}^2.
\end{equation}
If we start from a wavepacket in the origin, with momentum $p_0$, it will undergo harmonic oscillations. Their frequency is renormalized,
\begin{equation}
\Omega_{\rm p} = \Omega_{\rm I} \sqrt{M/M_{\rm p}},
\label{eq:FreqRen}
\end{equation}
due to the enhanced polaron mass.

The amplitude of the harmonic oscillations, $\sigma$, is easily obtained from energy conservation, $p_0^2/2M_{\rm p} = M \Omega_{\rm I}^2 \sigma^2/2$. By the same polaronic mass enhancement, this amplitude is also renormalized compared to the case of a free impurity,
\begin{equation}
\sigma = \sqrt{\frac{M}{M_{\rm p}}} \frac{p_0}{M \Omega_{\rm I}}.
\label{eq:AmpRen}
\end{equation}

Now we will use full dynamical simulations to show that both the frequency and the amplitude renormalization of polaron oscillations can serve as indicators of the effective polaron mass. In the case of a homogeneous Bose gas, Eqs.~\eqref{eq:FreqRen}, \eqref{eq:AmpRen} provide an accurate description of polaron trajectories at weak couplings.

\begin{figure}[b!]
\centering
\epsfig{file=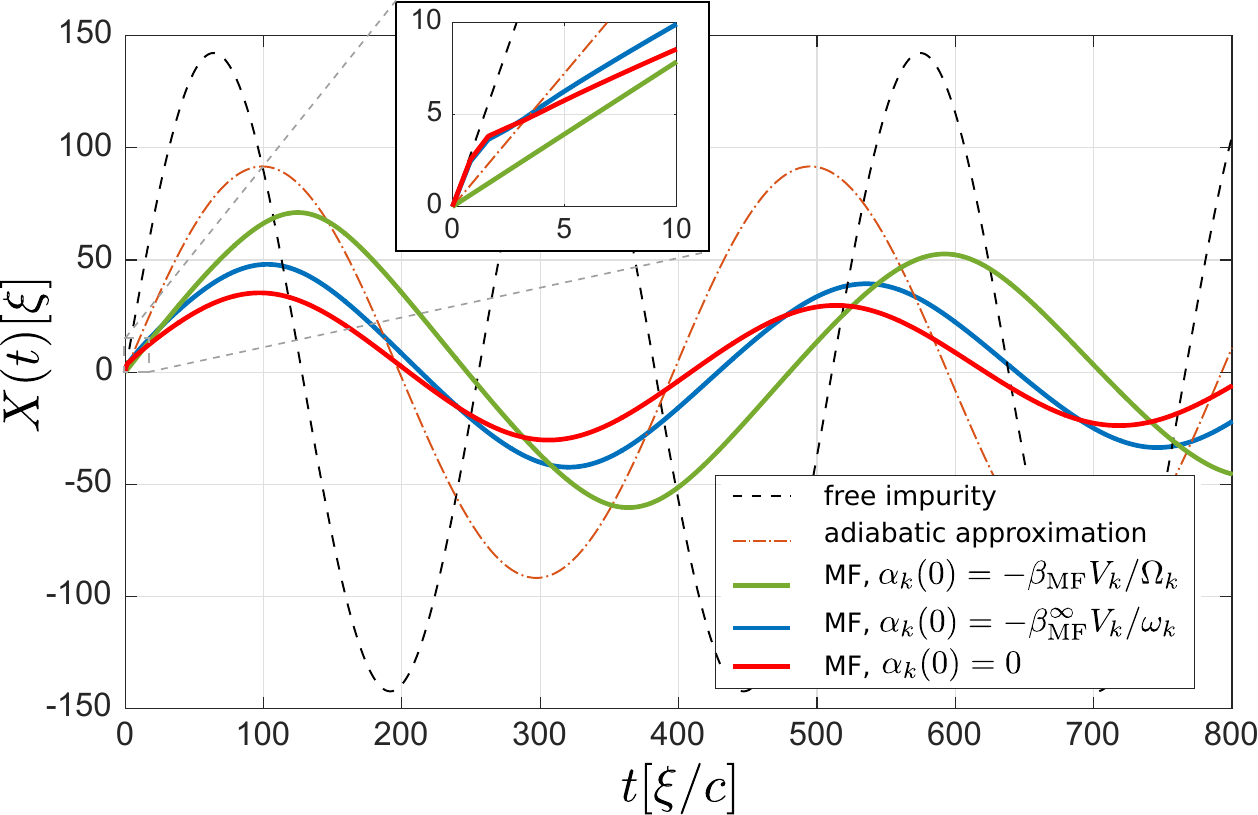, width=0.48\textwidth}
\caption{Typical polaron trajectories are shown for different initial conditions of the phonon cloud around the impurity. We used a trapping potential $\Omega_{\rm I} = 0.025 c/\xi$ as in Ref.~\cite{catani2012quantum}, which we treated in LDA. The Bose gas was assumed to be homogeneous in this case and the initial momentum $p_0= 3.5 M c$ of all trajectories corresponds to the typical thermal energy $k_{\rm B} T = p_0^2/2M$ in the experiment by Catani et al.~\cite{catani2012quantum}. The impurity-boson interaction was $\eta=4$, and the remaining parameters were chosen as in Ref.~\cite{catani2012quantum}. We used a UV cut-off $\Lambda_0=3/\xi$ and checked that the results have converged.}
\label{fig:polaronOscillationsInitials}
\end{figure}

\begin{figure*}[t!]
\centering
\epsfig{file=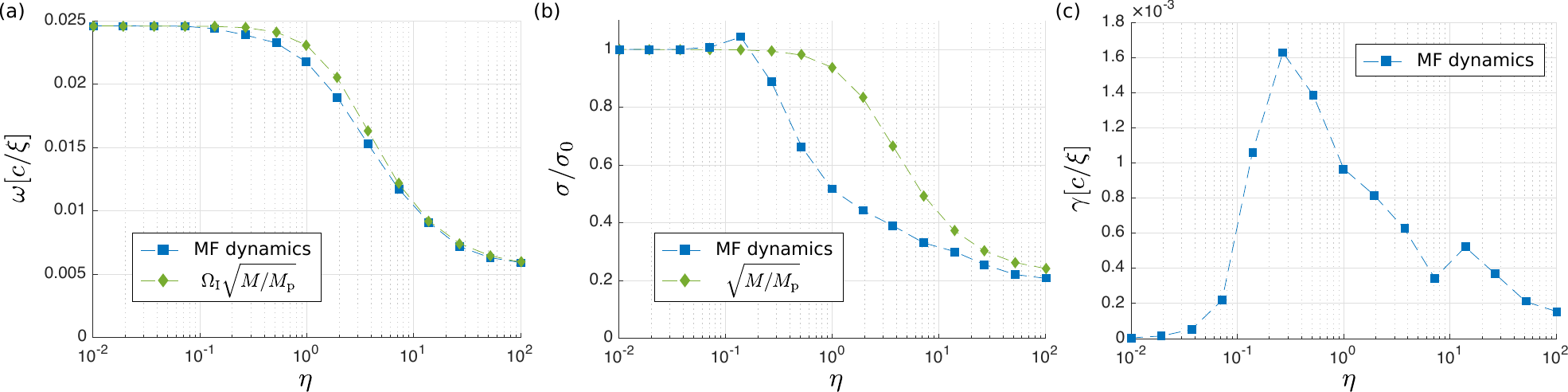, width=\textwidth}
\caption{Polaron oscillations in a homogeneous Bose gas: We simulate impurity trajectories as expected in a situation similar to the experiment by Catani et al.~\cite{catani2012quantum}. To this end we average over wavepackets, which we treat in LDA, with a momentum distribution determined by the temperature $k_{\rm B} T \approx 2 c/\xi$. The resulting trajectories $\sigma(t)$ are fitted to Eq.~\eqref{eq:sigmatFit}. Here we plot the fitting parameters for the frequency $\omega$ (a), the amplitude $\sigma$ (b) and the damping rate $\gamma$ of the resulting polaron oscillations as a function of the coupling strength. Except for the assumption of a homogeneous Bose gas we used parameters to describe the situation in Ref.~\cite{catani2012quantum}. To facilitate the dynamical simulations, we chose a soft UV cut-off $W_k \to W_k e^{-4k^2/\Lambda_0^2}$ and used $\Lambda_0=3/\xi$.}
\label{fig:polaronOscillationsResults}
\end{figure*}

\begin{figure*}[t!]
\centering
\epsfig{file=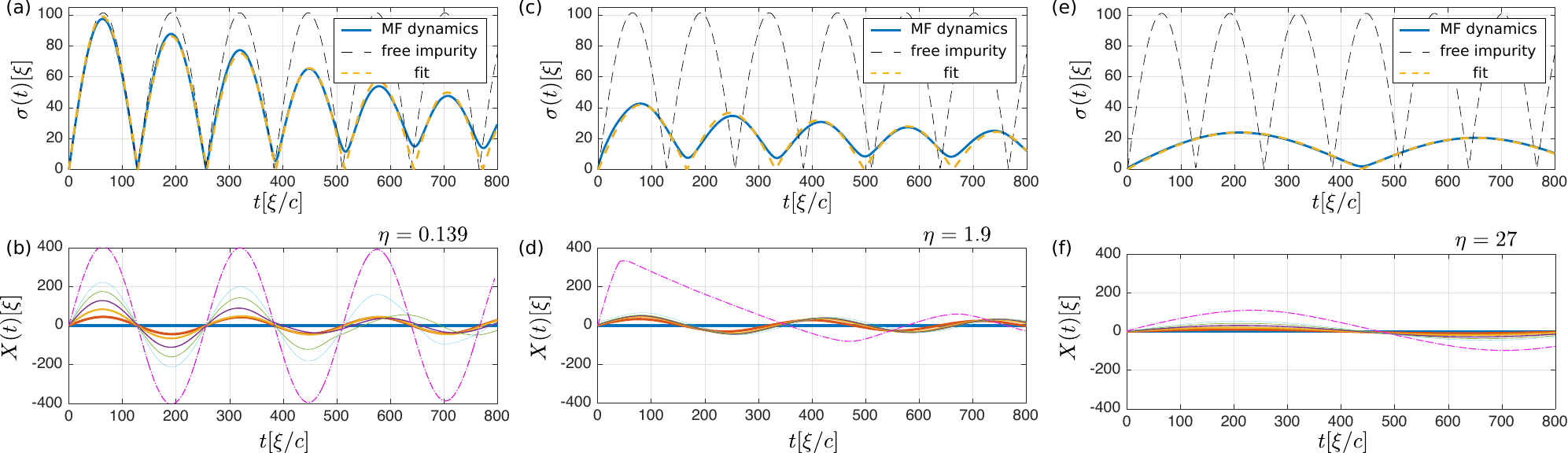, width=\textwidth}
\caption{Polaron oscillations inside a homogeneous Bose gas are simulated using the LDA. The top row shows the expansion of the impurity cloud $\sigma(t)$, see Eq.~\eqref{eq:sigmatDef}, overaged over initial polaron momenta, along with the fit according to Eq.~\eqref{eq:sigmatFit} and the case of a free impurity. In the bottom row trajectories in the ensemble are shown, with a width corresponding to their thermal weight and starting only with positive momenta. The trajectories with the largest initial velocity are highlighted by the dash-dotted lines. In the weak coupling regime (a), (b) for $\eta = 0.139$ we observe long-lived but decaying oscillations. For intermediate interactions (c), (d) at $\eta=1.9$ the velocity of almost all trajectories quickly drops below the speed of sound $c$. This happens almost instantly in the strong coupling regime (e), (f) where $\eta=27$. Parameters are the same as in Fig.~\ref{fig:polaronOscillationsResults}.}
\label{fig:polaronOscillationsTrajectories}
\end{figure*}

\emph{Initial conditions.--}
We use initial conditions as close as possible to the experimental situation described in Ref.~\cite{catani2012quantum}. There the impurities were tightly confined in the species-selective dipole trap, before they were released and their breathing oscillations were recorded. To model this situation, we consider polaron wavepackets localized in the origin $x=0$. We assume that the distribution $f(p_0)$ of their momenta $p_0 = P(0)$ can be derived from the thermal state of the impurity in the initial, tight trapping potential. Then we perform numerical calculations for different initial system momenta $P(0) = p_0$ and add up the resulting trajectories with their respective thermal weights.

We also need the initial phonon configuration in the polaron frame. Because the species-selective dipole trap provides a rather deep trapping potential, $\omega_{\rm I} = 0.28 c /\xi$, we think it is reasonable to assume that the impurity is initially localized, corresponding to $M=\infty$. Therefore the phonon cloud at time $t=0$ is chosen as the MF solution at $M =\infty$,
\begin{equation}
\alpha_k(0) = - \beta^\infty_\MF \frac{V_k}{\omega_k}, \quad \beta^\infty_\MF = \l  1 + \frac{g_\IB}{2 \pi} \int dk \frac{W_k^2}{\omega_k} \r^{-1}.
\end{equation}

We study the influence of the initial phonon distribution on the long-time behavior in Fig.~\ref{fig:polaronOscillationsInitials}. We compared cases where the MF solution at finite mass $M$, infinite mass $M=\infty$, or the phonon vacuum $\alpha_k=0$, was chosen initially. In Fig.~\ref{fig:polaronOscillationsInitials} we show typical polaron trajectories calculated within LDA for moderate interactions. Unless the MF solution at finite $M>0$ is chosen as initial state, the polaron loses a substantial part of its momentum at short times (see inset), until its velocity drops below the speed of sound $c$. This dissipation leads to a reduced oscillation amplitude, which manifests itself in the trajectories even at long times.

\emph{Numerical results.--}
To study polaron oscillations systematically, we simulated the experiment by Catani et al.~\cite{catani2012quantum} but assumed a homogeneous Bose gas first. In Fig.~\ref{fig:polaronOscillationsInitials} a typical polaron trajectory is shown, starting from an initial momentum $P(0)$ which is characteristic for the temperatures in the experiment. Note that it corresponds to a supersonic impurity with velocity $\dot{X}(0)>c$. At long times we observe long-lived but decaying oscillations with polaron velocities which are always below the speed of sound $c$.

We repeated these calculations for different interactions $\eta$ and averaged over the thermal distribution of initial momenta $P(0)$ corresponding to Ref.~\cite{catani2012quantum}. Then the resulting trajectories
\begin{equation}
\sigma(t) = \sqrt{ \langle X^2(t) \rangle_T},
\label{eq:sigmatDef}
\end{equation}
where $\langle \cdot \rangle_T$ corresponds to thermal averaging, were fitted to a function,
\begin{equation}
\sigma(t) = \sigma |\sin (\omega t)| e^{- \gamma t}.
\label{eq:sigmatFit}
\end{equation}
In Fig.~\ref{fig:polaronOscillationsResults} the fit parameters $\sigma$, $\omega$ and $\gamma$ are shown as a function of the interaction strength $\eta$. Some of the trajectories from which these plots were derived are plotted in Fig.~\ref{fig:polaronOscillationsTrajectories}.

In Fig.~\ref{fig:polaronOscillationsResults}(a) we compare results for the frequency renormalization with predictions by the adiabatic approximation, see Eq.~\eqref{eq:FreqRen}, and obtain excellent agreement. Our findings demonstrate that a frequency measurement of polaron oscillations provides an accurate method to detect the effective polaron mass. In particular, it also works in the strongly interacting regime and shows the saturation of the polaron mass in this case.

In Fig.~\ref{fig:polaronOscillationsResults}(b) we compare results for the amplitude renormalization to predictions by the adiabatic approximation, see Eq.~\eqref{eq:AmpRen}. The amplitude $\sigma_0$ is determined by the temperature of the impurity in this case; It is on the order of $\sigma_0 \approx \Omega_{\rm I}^{-1} \sqrt{2 k_{\rm B} T/M}$. Both for weak and strong impurity-boson interactions we find that the amplitude of impurity oscillations is directly related to the effective polaron mass according to the adiabatic expression in Eq.~\eqref{eq:AmpRen}. In the intermediate regime large deviations from this prediction can be observed.

To understand this behavior we take a closer look at the impurity trajectories. In Fig.~\ref{fig:polaronOscillationsTrajectories} (b) we observe that, for small interactions, the impurity oscillates back and forth several times at supersonic speeds. It is slowed down continuously until eventually it becomes subsonic. In this regime the amplitude of oscillations is well approximated by the adiabatic expression~\eqref{eq:AmpRen}. At intermediate interactions the impurity quickly loses kinetic energy and becomes subsonic, see Fig.~\ref{fig:polaronOscillationsInitials} and \ref{fig:polaronOscillationsTrajectories}(d). This dissipative effect goes beyond the adiabatic approximation and explains the strongly enhanced amplitude renormalization found in Fig.~\ref{fig:polaronOscillationsResults}(b) for this regime. For strong interactions, the impurity velocity drops below the speed of sound almost instantly for all initial velocities, see Fig.~\ref{fig:polaronOscillationsTrajectories}~(f). This suggests that the initial impurity momentum is almost reversibly transformed into polaron momentum in this regime. Afterwards the propagation of the heavy polaron can be described by the adiabatic approximation again.

In Fig.~\ref{fig:polaronOscillationsResults}(c) we present results for the damping rate of polaron oscillations in a harmonic trap. As before, three regimes of weak, intermediate and strong interactions can be identified. In agreement with our previous analysis, the decay is largest for moderate interactions. In this regime the non-adiabatic corrections are maximal: for stronger coupling the impurity can follow the phonon cloud more easily, suppressing non-adiabatic processes. For weaker couplings the probability for a non-adiabatic process where a phonon is emitted is strongly reduced because it scales as $\eta^2$.

\emph{Relation to the Florence experiment.--}
In this section we neglect the additional complication arising from an inhomogeneous Bose gas and compare experimental results to the theoretical analysis that assumes uniform boson density. We discuss the role of inhomogeneity in the following section \ref{subsec:polaronOsc2}.

Our analysis in Fig.~\ref{fig:polaronOscillationsResults} suggests that the oscillation frequency is the most useful observable for measurement of the effective polaron mass. Surprisingly no frequency renormalization has been found by Catani et al.~\cite{catani2012quantum}. On the other hand, the amplitude renormalization has been observed and as shown in Fig.~\ref{fig:1DpolaronMass} its comparison with our different theoretical calculations is excellent at weak couplings. At larger interactions the agreement is not as good, and the deviations are opposite from what we would have expected according to Fig.~\ref{fig:polaronOscillationsResults}(b).

From Fig.~\ref{fig:polaronOscillationsResults}(c) we read off a typical damping rate $\gamma \approx 10^{-3} c/\xi$ around $\eta=1$. The breathing frequency $\omega_b$, which is twice the oscillation frequency $\omega$, is on the order of $\omega_b=0.04 c/\xi$ in this regime. Therefore we expect a friction coefficient $\tilde{\gamma} := \gamma / 2 \omega_b$, as introduced in Ref.~\cite{catani2012quantum}, of $\tilde{\gamma} \approx 0.025$. This value agrees well with the measured value of $\tilde{\gamma} =0.03(2)$ for this interaction strength.

\subsection{Polaron oscillations: inhomogeneous Bose gas}
\label{subsec:polaronOsc2}
As we discussed above, the experimental observations in Ref.~\cite{catani2012quantum} are not consistent with our predictions for polaron oscillations inside a homogeneous Bose gas. Now we discuss possibilities how the inhomogeneity of the Bose gas can affect the oscillations of the impurity.

First we start by including the effective polaronic potential from Eq.~\eqref{eq:Vpolx} in the simulations. Some comments are in order. The use of this potential can only be justified in the limit where the polaron cloud follows the impurity adiabatically. Moreover, the treatment becomes meaningless close to the boundary of the Bose gas. In this regime the boson density is small and the Bogoliubov approximation breaks down. Moreover the Thomas-Fermi approximation is not accurate in this regime. Therefore the following results should be considered as a qualitative guide to understanding experiments rather than accurate quantitative analysis. 

\begin{figure}[t!]
\centering
$\quad$ \epsfig{file=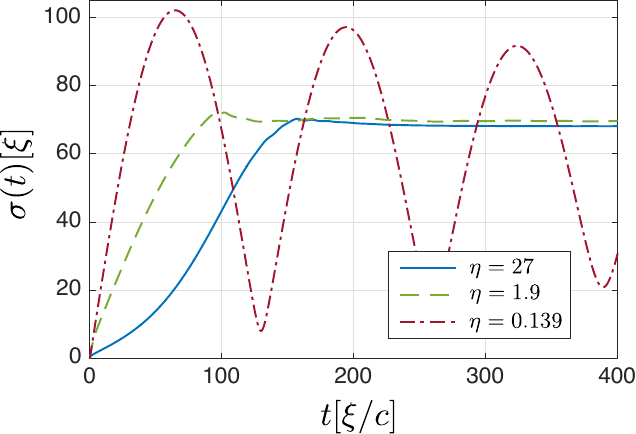, width=0.4\textwidth} $\qquad$
\caption{Polaron trajectories $\sigma(t)$ for an impurity described by the effective polaronic potential in Eq.~\eqref{eq:Vpolx}. Calculations include inhomogeneous density of the Bose gas in a trap and averaging over thermal distribution of the initial momentum of the impurity. We used the same parameters as in Fig.~\ref{fig:polaronOscillationsResults}, except that the potential $\Omega_{\rm B} = 0.0175 c / \xi$ was included. The Thomas-Fermi radius of the quasi-1D condensate in this case is $R\approx 80 \xi$.}
\label{fig:polaronOscillationsInTrap}
\end{figure}

The polaron trajectories in the presence of the polaronic trapping potential are shown in Fig.~\ref{fig:polaronOscillationsInTrap}. For weak impurity-boson interactions, the result is similar to the case of a homogeneous Bose gas and we obtain long-lived oscillations, slightly renormalized compared to the free impurity. For larger interactions all trajectories quickly become localized at the boundary of the Bose gas. Similar behavior has been reported in Ref.~\cite{catani2012quantum} for long times, although pronounced oscillations have also been seen in that case.

We now discuss that absence of frequency renormalization observed in Ref.~\cite{catani2012quantum} in the strongly interacting regime may be a result of inhomogeneous density of bosons. Let us consider an impurity trajectory with an amplitude which exceeds the size of the Bose gas by a sizable amount. Moreover, when the particle is interacting with the bosons, let us apply the adiabatic approximation. We may thus assume that all the initial momentum of the impurity is adiabatically transferred into polaron momentum. Indeed, we have shown for the case of a homogeneous gas that this simple picture explains why the amplitude of polaron oscillations is renormalized by a factor of $\sqrt{M/M_{\rm p}}$ for very strong couplings. This argument only relies on the conservation of energy, and therefore it also applies when the impurity exists the region where the Bose gas is confined.

For sufficiently large interactions, Fig.~\ref{fig:EffPotential} shows that the effective polaronic potential can be steeper than the harmonic confining potential for the impurity. Therefore it is conceivable that the impurity spends more time outside the Bose gas, than inside. In an extreme case it may even be reflected off the boundary. In such a situation the breathing frequency $\omega_b$ is given by twice the bare trap frequency, $\omega_b = 2 \Omega_{\rm I}$, as observed in Ref.~\cite{catani2012quantum}. The oscillation amplitude, on the other hand, is determined by the initial energy of the impurity inside the Bose gas, which can be related to the effective polaron mass.

\section{Beyond the Bogoliubov approximation}
\label{sec:BeyondBogoliubov}
So far we mostly ignored the effects of interactions between phonons. There are at least two good reasons to include them in our discussion and investigate the full Hamiltonian,
\begin{equation*}
\H = \H_F + \H_{\rm 2ph} + \H_{\rm ph-ph}.
\end{equation*}
First of all, the experiment by Catani et al.~\cite{catani2012quantum} is in a rather strongly interacting regime. The LHY corrections to the ground state energy, see Eq.~\eqref{eq:defLHY}, are on the order of $30 \%$ in this case. This indicates that the Bogoliubov approximation may not be justified for describing the experimental observations.

Secondly, we are interested in regimes where the impurity-boson coupling becomes dominant. As we have shown, this leads to an accumulation of a large number of phonons (bosons) around the impurity, see e.g. Fig.~\ref{fig:depletion}. Even if the boson interaction strength $g_{\rm BB} \ll g_\IB$ is negligible compared to the impurity-boson coupling, the interaction energy $\sim g_{\rm BB} n(x)^2$ can become comparable to the polaron energy $\sim g_\IB n(x)$.
This is possible because of the quadratic scaling of the boson's interaction energy with their density $n(x)$ around the impurity. The break-down of the RG (see Appendix \ref{appdx:RGapproach}) also indicates that interactions between the phonons have to be included to prevent the system from becoming unstable to quantum fluctuations.

In this section we explain the DMC method (\ref{subsec:DMCapproach}) which we used for calculating the ground state energy of the full Hamiltonian $\H$, as well as the effective mass of the impurity. We benchmark this method (in \ref{subsec:GPElimit}) by discussing the exactly solvable case of an impenetrable impurity inside a weakly interacting Bose gas. This leads us to a final discussion of the experiment by Catani et al.~\cite{catani2012quantum} (in \ref{subsec:compCataniEnergy}), where we focus on the effect of boson-boson interactions.

\subsection{The DMC method for Bose polarons}
\label{subsec:DMCapproach}

Monte Carlo methods provide an efficient tool for the evaluation of multidimensional integrals.
Expectation values of quantum mechanical operators can be written as integrals over the ground state wave function and can be efficiently evaluated using Monte Carlo techniques.
Here we resort to the diffusion Monte Carlo method in order to obtain the ground state properties of the Bose polaron problem. For a general reference on the DMC method, see for example~\cite{Boronat1994}. An important advantage of the non-perturbative DMC method is that it can be applied to arbitrary interaction strength $\eta$ between the impurity and the surrounding Bose gas. This is true for arbitrary values of the gas parameter $\gamma$ of the bath, see Eq.~\eqref{eq:defGamma}.

Another question which can be addressed by performing DMC simulations concerns the dimensionality of the system. Specifically, we study the effect of the transverse confinement potential in the Florence experiment~\cite{catani2012quantum} on the energetic and dynamical properties of the system. To this end we consider both the effective 1D Hamiltonian with contact interactions, see Eq.~\eqref{eq:Hmicro}, and a full 3D Hamiltonian. The latter takes into account the geometry of the optical lattice potential as well as the three-dimensional scattering length. Even though the experimental geometry is rather complicated, with different optical lattice potentials felt by the Rb and K atoms, the DMC method can still be applied.

We focus on two densities, first corresponding to strong interactions in the bath, $n_0|a_{\rm BB}| = 4.56$ or $\gamma=0.44$, as in the Florence experiment~\cite{catani2012quantum}.
The second considered value, $n_0|a_{\rm BB}| = 144$, corresponds to $\gamma=0.014$ deep in the Bogoliubov regime.
For the first case we consider impurities (Rb) and bath particles (K) of different mass.
In the second case we will also consider the limit of an infinitely repulsive pinned impurity, in which Bethe ansatz can be used in order to find the ground state energy of the Bose polaron.
This permits us to verify the consistency of the DMC energy with the Bethe ansatz result in this exactly integrable limit. In addition we obtain the correlation functions and the density profile of bosons around the impurity from DMC calculations.

\emph{Three-dimensional DMC calculations.--} In first quantization the 3D model Hamiltonian is
\begin{widetext}
\begin{equation}
\H = -\frac{\hbar^2}{2M}\Delta_I
-\frac{\hbar^2}{2m_{\rm B}}\sum\limits_{i=1}^N\Delta_i  +\sum\limits_{i<j}^N V_{\rm BB}({\bf r}_i - {\bf r}_j)
+\sum\limits_{i=1}^N V_{\rm IB}({\bf r}_i - {\bf r}_I)
+V^{{\rm ext}}_{\rm I}({\bf r}_I)+\sum\limits_{i=1}^NV^{\rm ext}_{\rm B}({\bf r}_i).
\label{eq:H3D}
\end{equation}
\end{widetext}
Here $V_{\rm BB}(r)$ and $V_{\rm IB}(r)$ are the boson-boson and impurity-boson interaction potentials, and $\Delta_{I}$, $\Delta_i$ denote the Laplace operators with respect to the impurity and the boson labeled by $i$, respectively. The external potentials, $V^{\rm ext}_{\rm B}$ and $V^{\rm ext}_{\rm I}$, are felt by bosons and the impurity respectively. In the experiment \cite{catani2012quantum} they have been created by two-dimensional lattices forming an array of 1D tubes. We consider the case when a single tube in the array is populated.

The depth of the Rb lattice potential $V_{\rm latt}(r)$ is $s=60$ recoils with the lattice wavelength $\lambda = 1064$ nm,
$V_{\rm B}(r) = s \hbar^2/(2m_{\rm Rb}\lambda^2)$.
We have checked that similar result can be obtained by considering a simple harmonic external potential, $V_{\rm osc}(r)$, with the transverse oscillator frequency $\omega_\perp /2\pi = 34(45)$ kHz for Rb(K) atoms.
We ignore the residual shallow trapping along the longitudinal direction in DMC calculations.

The relation of the three-dimensional $s$-wave scattering length, $a_{\rm 3D}$, to the one-dimensional one, $a_{\rm 1D}$, for the tight transverse confinement with oscillator length $a_{\rm osc}$, is given by Olshanii's formula $a_{\rm 1D} = -a_{\rm osc}^2/a_{\rm 3D}(1-1.0326 a_{\rm 3D}/a_{\rm osc})$ \cite{Olshanii1998}.
In an optical lattice geometry no simple analytical result is known and the corresponding relation is obtained following \cite{Peano2005}.
In the Florence experiment \cite{catani2012quantum} the boson-boson $s$-wave scattering length was fixed to $a_{\rm 1D,Rb} = -652.2$nm and was not changed.
The boson-impurity $s$-wave scattering length, in contrast, is tunable over a wide range by changing the strength of the applied magnetic field.

In our quasi-1D simulations we model the three-dimensional interaction potential by hard-spheres, $V_{\rm HS}({\bf r}) = +\infty$ when $|{\bf r}|<a_{\rm HS}$ and $V_{\rm HS}({\bf r}) =0$ otherwise. The diameter of the hard-sphere potential coincides with its $s$-wave scattering length and is set to reproduce Rb-Rb (Rb-K) value for the boson-boson (boson-impurity) scattering amplitude. In our simulations we consider a single impurity and impose periodic boundary conditions along the longitudinal direction of the tube.

The statistical fluctuations in Monte Carlo simulation can be greatly reduced by using importance sampling. This is done on the basis of a distribution function which we derive from a trial guiding wave function. Motivated by our physical insights, it is chosen as a product of one- and two-body terms,
\begin{eqnarray}
\nonumber
\psi_T({\bf r}_1, \cdots, {\bf r}_N; {\bf r}_I)
=
h_{\rm I}({\bf r}_I)\prod\limits_{i=1}^N h_{\rm B}({\bf r}_i)\\
\prod\limits_{i<j}^N f_{\rm BB}(|{\bf r}_i - {\bf r}_j|)
\prod\limits_{i=1}^N f_{\rm IB}(|{\bf r}_i - {\bf r}_I|).
\label{wf:3D}
\end{eqnarray}
The Gaussian one-body terms $h_{\rm I}({\bf r}) = \exp[-(x^2+y^2)/2a_{\rm osc,K}^2]$ and $h_{\rm B}({\bf r}) = \exp[-(x^2+y^2)/2a_{\rm osc,Rb}^2]$ localize the particles inside the central tube. The two-body Jastrow terms are taken in the following form
\begin{eqnarray}
f_{\alpha}(r) =
\begin{cases}
0, r<a_{\alpha}\\
\frac{A}{r}\sin(B(r-a_{\alpha}))\\
1, r>L/2
\end{cases}
\label{wf:Jastrow}
\end{eqnarray}
where parameters $A$ and $B$ are chosen such that both $f_{\alpha}(r)$ and its derivative are continuous at the half size of the box, $r=L/2$.
Here $\alpha={\rm BB}$ corresponds to boson-boson and $\alpha={\rm IB}$ to the case of impurity-boson scattering. The Jastrow terms~(\ref{wf:Jastrow}) are obtained as the solutions of the two-body scattering problem on the hard sphere potential.
Although the guiding wave function~\eqref{wf:3D} - \eqref{wf:Jastrow} does not contain any variational parameters, it provides a sufficient quality.

\emph{One-dimensional DMC calculations.--}
We also perform calculations for the 1D Hamiltonian from Eq.~\eqref{eq:Hmicro}. In this case the guiding wave function can be conveniently written in a pair-product form
\begin{eqnarray}
\nonumber
\psi_T(x_1, \cdots, x_N; x_I)
=
\prod\limits_{i<j}^N f_{\rm BB}(|x_i - x_j|)
\prod\limits_{i=1}^N f_{\rm IB}(|x_i - x_I|).
\label{wf:1D}
\end{eqnarray}
We chose the two-body terms as
\begin{eqnarray}
f_\alpha(x) =
\begin{cases}
A_\alpha \cos(k_\alpha(x-B_\alpha)), \quad  |x|\leq R^{\rm par}_\alpha\\
|\sin(\pi x/L)|^{1/K_\alpha^{\rm par}}, \qquad R^{\rm par}_\alpha<|x|
\end{cases}
\label{wf:Jastrow1D}
\end{eqnarray}
where $\alpha = $ BB, BI denotes boson-boson (boson-impurity) terms.
Parameters $A_\alpha$ and $B_\alpha$ are chosen in such a way that $f_\alpha(x)$ and its first derivative are continuous for $|x|>0$ and satisfy Bethe-Peierls boundary condition at the contact, $f_\alpha'(0)=-f_\alpha(0)/a_\alpha$. Here $a_\alpha$ denotes the corresponding 1D $s$-wave scattering length. The short-range part in Eq.~(\ref{wf:Jastrow1D}), for $|x|\leq R^{\rm par}_\alpha$, is the two-body scattering solution for a contact $\delta$-function potential. Instead the ``phononic'' long-range part in Eq.~(\ref{wf:Jastrow1D}), for $ |x| > R^{\rm par}_\alpha$, is obtained from the hydrodynamic approach \cite{Reatto1967}.

The variational parameter $R^{\rm par}_\alpha$ corresponds to the crossover distance between the two-body and phononic regimes. It is optimized by minimizing the variational energy, which leads to the variational Monte Carlo (VMC) results presented earlier in this paper. The variation parameter $K^{\rm par}_{\rm BB}$ for large system size coincides with the Luttinger parameter of the bath $K$. Its dependence on the gas parameter, $K( \gamma )$, is known from the Bethe ansatz solution to the Lieb-Liniger model \cite{Lieb1961}.
We use the thermodynamic value of parameter $K^{\rm par}_{\rm BB}$ for the bath and optimize parameter $K^{\rm par}_{\rm IB}$ by minimizing the variational energy.

\emph{Variational Monte Carlo method.--}
The variational Monte Carlo method evaluates averages over the trial wave function $\psi_T$. The Metropolis algorithm \cite{METROPOLIS1953} is used to sample its square, $|\psi_T|^2$ by generating a Markov chain with corresponding probability distribution. The average of the Hamiltonian, $E_{\rm var} = \bra{\psi_T}\H\ket{\psi_T} / \bra{\psi_T} \psi_T \rangle$, provides an upper bound to the ground state energy $E_0$. It is interesting to compare the value of $E_{\rm var}$ with prediction of MF theory.

\emph{Diffusion Monte Carlo method.--}
The diffusion Monte Carlo method~\cite{Boronat1994} is based on solving the Schr\"odinger equation in imaginary time.
For large times, the contribution to the energy from the excited states is exponentially suppressed, permitting to obtain the exact ground state energy $E_0$.
The density profile of the polaron can be calculated using the technique of pure estimators \cite{Liu1974,Sarsa2002}.
The effective mass of the polaron, $M_{\rm p}$, is obtained by calculating the diffusion coefficient $D$ of the impurity in the imaginary time $\tau$,
$D = \lim_{\tau\to\infty} \langle [r_{\rm I}(\tau) - r_{\rm I}(0)]^2\rangle /\tau$,
according to the relation $D = \hbar^2/2M_{\rm p}$.
The variational value for $M_{\rm p}$ is obtained from DMC algorithm without branching, which is an alternative method to the Metropolis algorithm to generate the probability distribution according to $|\psi_T|^2$.

\subsection{The GPE limit: Dark solitons}
\label{subsec:GPElimit}
A physically important limit is that of an impurity with an infinite mass, $M = \infty$. It corresponds to two realistic situations: (i) a pinned impurity, (ii) a static potential created by a focused laser beam. Furthermore this limit is interesting as it is allows to obtain physical insights to the effects of phonon-phonon interactions on the polaron cloud.

When the Bogoliubov approximation is justified -- which is the case for $\gamma \ll 1$ or $n_0 |a_{\rm BB}| \gg 1$ -- we can study this situation using the Gross-Pitaevskii mean-field equation (GPE)~\cite{Pitaevskii2003}:
\begin{equation}
E_{\rm B} \phi(x) = \left[ - \frac{\partial_x^2}{2 m_{\rm B}} + g_{\rm BB} |\phi(x)|^2 + g_{\rm IB} \delta(x) \right] \phi(x).
\label{eq:GPE}
\end{equation}
Here $\phi(x) = \langle \hat{\phi}(x) \rangle$ describes the boson field in the 1D system, and $E_{\rm B}$ denotes the total energy $E_{\rm B}$ of the combined impurity-boson system.

The GPE is valid even for strong impurity-boson interactions and goes beyond perturbative expansions in orders of $g_{\rm IB}$. This is different from perturbative theories which linearize boson-impurity interactions, which is justified when $g_{\rm IB}$ is small. Because the boson-boson interactions are treated on a mean-field level, the validity of GPE is limited to the Bogoliubov regime $n_0 |a_{\rm BB}| \gg 1$ ($\gamma \ll 1$) as we discuss below.

\emph{Dark soliton solution.--}
The repulsive non-linear GPE possesses a famous class of solutions known as gray solitons. These correspond to a depletion in the boson density which maintains its shape while propagating with a constant speed. In the case of zero velocity the density completely vanishes in a single point, and the solution is referred to a a dark soliton. Thus we expect that the effect of a massive impenetrable impurity, with $M, g_\IB \to \infty$, is to localize a dark soliton in the Bose gas.

The wave function of a dark soliton is given by $\phi_0(x) = \sqrt{n_0} \tanh ( x / 2 \xi )$ with the corresponding energy $E_{\rm B} = g_{\rm BB} n_0 N/2 -2 n_0 c/3$.
If we take into account that $\Delta N_{\rm sol} = 2 \sqrt{2} n_0 \xi$ bosons are repelled from the homogeneous Bose gas, we find that the energy $E_0$ of the impurity in a system with fixed total particle number $N$ and density $n_0$ is
\begin{equation}
E_0^{\rm sol} = \frac{4}{3} n_0 c.
\label{eq:e0Soliton}
\end{equation}
This energy has the same physical meaning as other polaron energies $E_0$ which we calculated before.

On the other hand, we can calculate the polaron energy starting from an impurity and using MF theory as described in Sec.~\ref{subsec:MFtwoPhonon}. In that case we obtain
\begin{equation}
E_0^\MF = 2 n_0 c = \frac{3}{2} E_0^{\rm sol}
\label{eq:e0MFcfSoliton}
\end{equation}
by sending $g_\IB \to \infty$ and $M \to \infty$. The obtained expression overestimates the correct energy by the factor of $3/2$. The reason is that in our MF theory we ignore the non-linearity in Eq.~\eqref{eq:GPE}. By making the Bogoliubov approximation we effectively linearize $|\phi|^2 \phi \approx n_0^{3/2} + 2 n_0 \delta \phi(x) + n_0 \delta \phi^*(x)$.

\begin{figure}[t!]
\centering
\epsfig{file=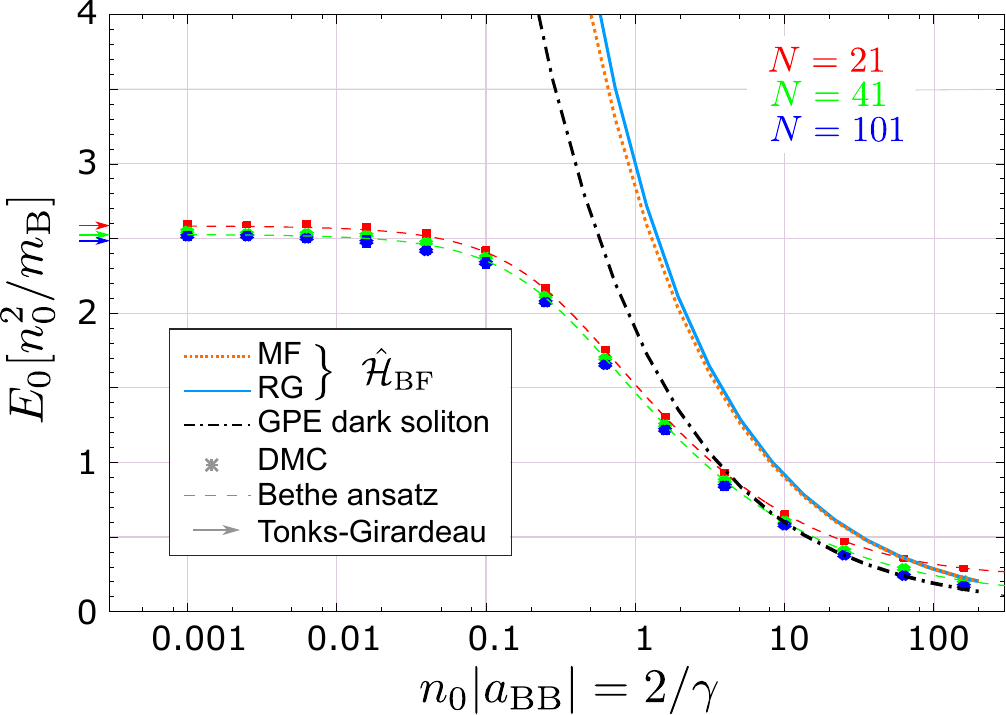, width=0.4\textwidth} $\qquad$
\caption{The energy shift $E_0$ caused by an impenetrable static impurity ($M=\infty$ and $g_\IB=\infty$) is calculated as a function of the interaction parameter $n_0 |a_{\rm BB}|$ of the surrounding Bose gas. While MF and RG rely on the Hamiltonian $\H_{\rm BF}$, the other approaches start from the full model $\H$. The dash-dotted line corresponds to the dark-soliton solution of the Gross-Pitaevskii equation (GPE). To ensure proper finite-size scaling, DMC results are compared to exact analytical calculations using Bethe ansatz and in the Tonks-Girardeau limit.}
\label{fig:pGPE}
\end{figure}

\emph{Comparison to DMC.--}
To benchmark our theoretical methods we calculated the energy of an impenetrable, localized impurity for different parameters of the Bose gas. Our comparison in Fig.~\ref{fig:pGPE} shows that DMC is in perfect agreement with the exact result in the limit where Bogoliubov theory is valid, $n_0 |a_{\rm BB}| \gg 1$ or $\gamma \ll 1$. As expected, MF theory and RG differ by the factor $2/3$ in this regime.

In Sec.~\ref{subsec:RGapproach} we suggested to use the relative size of LHY corrections to the ground state energy of the Bose gas as an indicator where the Bogoliubov theory can be applied to describe polarons. This leads to the condition that $n_0 |a_{\rm BB}| \gg 1$ (or $\gamma \ll 1$). Although in Fig.~\ref{fig:pGPE} we never obtain quantitative agreement of the Bogoliubov approximation with the numerically exact DMC results, we find that the qualitative behavior of the impurity energy $E_0$ is correctly described in this framework for $n_0 |a_{\rm BB}| \gg 1$ (or small $\gamma \ll 1$). In the opposite limit $n_0 |a_{\rm BB}| \lesssim 1$ (or $\gamma \gtrsim 1$) in contrast, the Bogoliubov description completely fails. This shows that the LHY corrections provide a reliable measure for the accuracy of the truncated description of the polaron cloud using only the beyond-Fr\"ohlich Hamiltonian $\H_{\rm BF}$.

\subsection{Comparison to Florence experiment}
\label{subsec:compCataniEnergy}
In Fig.~\ref{fig:polaronEnergyStrongCoupling} we compare the polaron energy for repulsive interactions $g_\IB>0$ and parameters as in the experiment of Ref.~\cite{catani2012quantum}. For weak-to-intermediate impurity-boson interactions, $|\eta| \lesssim 1$, where the Fr\"ohlich model is valid and the density modulation in the Bose gas is small, all theoretical predictions agree with each other. For stronger couplings we observe sizable quantitative differences. Nevertheless, the qualitative behavior of all results going beyond the Fr\"ohlich Hamiltonian is the same. The corrections of the RG to the MF results is pronounced at large couplings, but DMC predicts even smaller polaron energies in this regime.

\begin{figure}[t!]
\centering
\epsfig{file=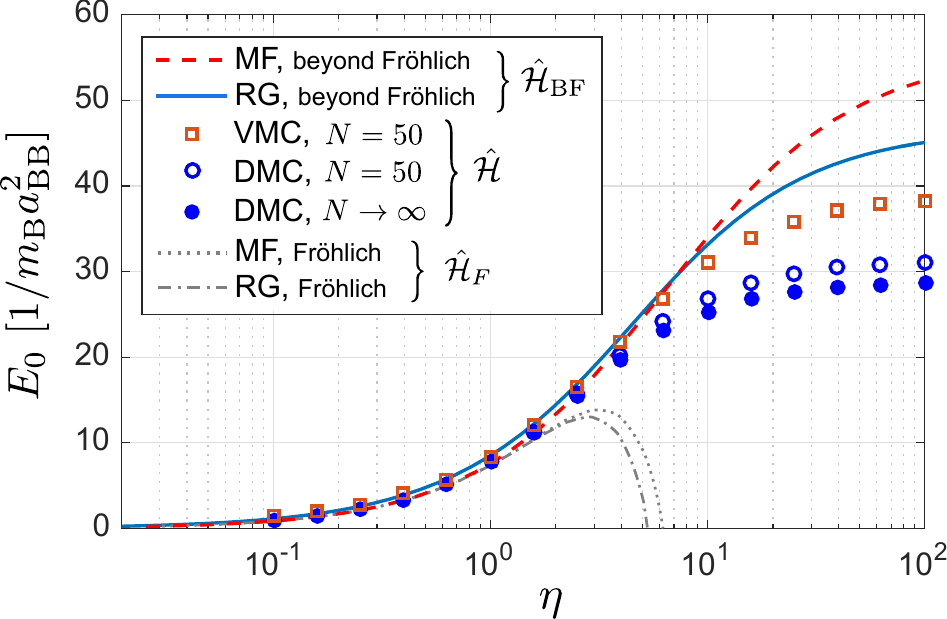, width=0.42\textwidth} $\qquad$
\caption{Comparison of the polaron energy $E_0$ computed by different methods and using Hamiltonians as indicated in the legend. We have chosen parameters as in the experiment by Catani et al.~\cite{catani2012quantum}. Note that the RG predicts a larger energy than MF theory for weak interactions because for the latter we ignored the logarithmically divergent term Eq.~\eqref{eq:e0MFlogDiv}. It is included and properly regularized in the RG. We show DMC and VMC results for $N=50$ particles. For DMC also the extrapolated value expected in the thermodynamic limit $N \to \infty$ is shown, see Appendix \ref{appdx:finite-size} for details of our analysis.}
\label{fig:polaronEnergyStrongCoupling}
\end{figure}

In Fig.~\ref{fig:1DpolaronMass} we compared predictions for the effective polaron mass to the experimental results from analyzing polaron oscillations. There we have found large deviations for $\eta \gtrsim 3$. Interestingly this is exactly where the strong coupling regime starts and DMC predicts different energies than our RG approach. In this regime the density modulations of the Bose gas around the impurity are expected to become large. As a result, our analysis of polaron oscillations showed that large deviations from the adiabatic result can be expected. We speculate that this may be related to the large difference between theory and experiment at strong couplings in Fig.~\ref{fig:1DpolaronMass}. 

In Ref.~\cite{catani2012quantum} it was moreover suggested that $\eta \approx 15$ could also be the point where higher transverse modes are important. Population of such modes would imply that the system can no longer be treated as strictly one-dimensional. However, the argument of Ref.~\cite{catani2012quantum} was based on a comparison of the \emph{bare} energy $g_\IB n_0$ with the transverse trapping frequency $\omega_\perp \approx 126 / m_{\rm B} a_{\rm BB}^2$. As can be seen from Fig.~\ref{fig:polaronEnergyStrongCoupling}, the relevant polaron energies are well below $\omega_\perp$ for all interaction strengths. Therefore we conclude that a cross-over into a higher-dimensional regime cannot explain the experimental observations. This supplements our analysis of higher transverse modes in Fig.~\ref{fig:1DphaseDiag}, where we arrived at the same conclusion.

We conclude by noting that phonon-phonon interactions play an important role for understanding polarons at strong couplings in the experiment by Catani et al.~\cite{catani2012quantum}. Not only do two-phonon terms beyond the Fr\"ohlich Hamiltonian become important, but also the non-linear interactions between phonons are required to describe correctly the polaron energy for repulsive interactions. On the attractive side, where $g_{\rm IB}<0$, we expect their influence to be even more dramatic, because larger deformations of the Bose gas around the impurity are possible.

\section{Summary and Outlook}
\label{sec:SummOut}

In this paper we studied theoretically mobile impurities interacting with a 1D Bose gas. We provided general theoretical analysis of such problems and considered a specific experimental system realized in experiments by Catani et al.~\cite{catani2012quantum}. We showed that in the weak coupling regime the Fr\"ohlich model provides an accurate description of the system. We extended our analysis to include two-phonon scattering terms, which become important for stronger impurity-boson interactions. Finally we also discussed the effects of boson-boson interactions on the polaron cloud.

\emph{Main new results.--}
The main new theoretical insights of our work are related to how two-phonon terms affect Bose polarons at strong coupling. Simple mean-field does not work in one dimension and needs to be corrected by RG calculations. For sufficiently weak boson-boson interactions we find that qualitative features of the polaron phase diagram remain the same as obtained from mean-field description of polarons in three dimensions~\cite{Shchadilova2016PRL,Grusdt2016RGBEC}. 

In particular we find a repulsive polaron branch for repulsive couplings, $\eta > 0$, and an attractive polaron branch for sufficiently weak attractive interactions, $\eta \lesssim 0$. In the weakly interacting limit, both branches can be described by the Fr\"ohlich polaron, which has been observed in the Florence experiment \cite{catani2012quantum}. On the other hand, for sufficiently strong attractive interactions, $\eta \ll 0$, we expect multi-particle bound states at low energies and a meta-stable repulsive polaron branch at high energies. The latter is adiabatically connected to the polaron at infinitely repulsive microscopic interactions $\eta \to \infty$. 

The enhanced role of quantum fluctuations in 1D manifests itself in the logartihmic divergence (with system size) of the MF polaron energy. In Appendix~\ref{appdx:RGapproach} we generalized the RG approach from Ref.~\cite{Grusdt2016RGBEC} to one dimension. We showed that the resulting polaron energy is regularized in the RG approach and converges to finite value when the system size is increased. Furthermore we showed that RG analysis can be used to study other properties of 1D Bose polarons, incudling the effective mass and impurity boson correlations. We compared these predictions to our numerically exact DMC calculations and found good agreement for a weakly interacting Bose gas in the Bogoliubov regime. We concluded that for a full quantitative description of Bose polarons at strong couplings, phonon-phonon interactions always need to be included. In addition we identified regimes in the phase diagram where the full microscopic Hamiltonian is required for reaching even a qualitative understanding of the polaron properties (Fig.~\ref{fig:Intro2}). 

\emph{Analysis of the experiment by Catani et al.~\cite{catani2012quantum}.--}
Original analysis of the experiments showed a disagreement between theoretical results and experimentally measured effective mass already for weak impurity-boson interaction. We explained that this disagreement results from the high energy modes with $k \sim 1/\xi$  that were not included properly in earlier analysis. We showed that both analytical RG and numerical DMC methods give results in good agreement with experiments when the high energy modes are included more accurately. 

Theoretical methods based on the Bogoliubov approximation predict a saturation of the polaron mass at a finite value of the impurity-boson interaction strength. Although qualitatively this behavior has been observed in the experiment, large quantitative deviations from our theoretical calculations are found in this regime, where effects beyond the Bogoliubov approximation are expected to play a role. We performed full numerical simulations of the polaron trajectories in a harmonic trapping potential and argued that the disagreement between theory and experiment could be related to the inhomogeneity of the Bose gas.

By performing DMC simulations for two different Hamiltonians:
(i) strictly one-dimensional, see Eq.~\eqref{eq:Hmicro},
(ii) three-dimensional, see Eq.~\eqref{eq:H3D}, with strong transverse confinement, we found no significant differences for the dynamic and static properties, even in the regime of strong interactions.
This means that the use of a strictly one-dimensional Hamiltonian is justified.

Closer inspection of different theoretical models revealed that in the strong coupling regime phonon-phonon interactions need to be included if one wants to do accurate comparison to experiments. In contrast to Bose polarons in three dimensions~\cite{Jorgensen2016PRL,Hu2016PRL}, the relative size of quantum fluctuation corrections to the ground state energy of the Bose gas was sizable in Ref.~\cite{catani2012quantum}. We showed here for 1D systems that this is a suitable indicator for the applicability of the Bogoliubov approximation for describing Bose polarons.

We conclude that a detailed quantitative analysis of the experiment by Catani et al.~\cite{catani2012quantum} at strong couplings is extremely challenging. We showed that it requires full inclusion of two-phonon terms $\H_{\rm 2ph}$ as well as phonon-phonon interactions $\H_{\rm ph-ph}$, both going beyond the Fr\"ohlich Hamiltonian that has been used previously to analyze the experiment \cite{catani2012quantum,Bonart2013}. Moreover, from full dynamical simulations of this problem we found indicators that the inhomogeneity of the Bose gas needs to be taken into account as well. We expect that numerical DMRG or TEBD \cite{White1993,White2004} calculations could shed new light on this problem in the future. Our analysis moreover ignored effects of finite temperatures, which can also contribute to the observed differences between theory and experiment.

\emph{Possible future experiments.--}
Here we performed full dynamical simulations of polaron trajectories inside a shallow trapping potential. We showed by using a time-dependent MF ansatz combined with the local-density approximation that polaron oscillations inside a homogeneous trap provide a powerful means for measuring the effective polaron mass. When the Bose gas can be assumed to be homogeneous, the frequency renormalization provides accurate results. We suggest to use species-selective optical traps in the future to perform such measurements, in a regime where the Bose gas is as large as possible to avoid effects of the inhomogeneous density profile.

The energy provides another important quantity to characterize Bose polarons. It can be obtained directly from the impurity's radio-frequency spectrum, which has been measured in 3D~\cite{Jorgensen2016PRL,Hu2016PRL}. We suggest to repeat these experiments in 1D systems, possibly even in the time-domain~\cite{Knap2012}. It would be particularly interesting to study quenches from strong repulsive to strong attractive interactions and show the existence of a repulsive polaron branch for attractive microscopic interactions $g_\IB < 0$ in one dimension.

\section*{Acknowledgements}
We acknowledge useful comments by Adrian Kantian and Thierry Giamarchi, as well as fruitful discussions with Yulia Shchadilova, Richard Schmidt, Thierry Giamarchi and Artur Widera. F.G. acknowledges support from the Gordon and Betty Moore foundation. G.E.A. acknowledges partial financial support from the MICINN (Spain) Grant No.~FIS2014-56257-C2-1-P. The authors acknowledge support from Harvard-MIT CUA, NSF Grant No. DMR-1308435, AFOSR Quantum Simulation MURI, AFOSR grant number FA9550-16-1-0323 and the Moore foundation. Authors thankfully acknowledge the computer resources at MareNostrum and the technical support provided by Barcelona Supercomputing Center (FI-2017-1-0009). The authors gratefully acknowledge the Gauss Centre for Supercomputing e.V. (www.gauss-centre.eu) for funding this project by providing computing time on the GCS Supercomputer SuperMUC at Leibniz Supercomputing Centre (LRZ, www.lrz.de). 

\appendix

\section{Mean-field theory for Bose polarons\\ in one dimension}
\label{sec:AppdxMF}
In this appendix we briefly derive the mean-field (MF) theory for Bose polarons in 1D. Our starting point is the beyond-Fr\"ohlich Hamiltonian $\H_{\rm BF}$ from Eq.~\eqref{eq:defBeyFr}, i.e. we apply the Bogoliubov approximation for describing the Bose gas and neglect interactions between the phonons. The standard MF theory developed for Fr\"ohlich polarons \cite{Lee1953,Devreese2013,BeiBing2009} has been generalized to this case in higher dimensions \cite{Shchadilova2016PRL}.

To solve the full Hamiltonian \eqref{eq:HFdef}, \eqref{eq:H2phDef} we start by applying the unitary Lee-Low-Pines (LLP) transformation \cite{Lee1953}. Following Refs.~\cite{Devreese2013,Grusdt2015Varenna,Shchadilova2016PRL} this gives rise to the Hamiltonian
\begin{widetext}
\begin{multline}
\H_{\rm LLP}(p) = g_\IB n_0 + \frac{1}{2 M} \l p - \int dk~ k \ad_k \a_k \r^2 + \int dk ~ \omega_k \ad_k \a_k + \int dk ~ V_k \l \ad_k + \a_{-k} \r +\\
+ \frac{g_{\rm IB}}{2 \pi} \int dk~ \sinh^2 \theta_k  + \frac{g_\IB}{2 \pi} \int d k ~ d k'~ : \l \cosh \theta_k \ad_{k} - \sinh \theta_k \a_{-k} \r \l \cosh \theta_{k'} \a_{k'} - \sinh \theta_{k'} \ad_{-k'} \r :,
\end{multline}
\end{widetext}
where $p$ is the total conserved system momentum. Note that we normal-ordered the two-phonon scattering terms (denoted by $:...:$), which gives rise to the constant energy shift in the first term of the second line.

In the MF theory of Bose polarons one makes an ansatz of coherent states $\prod_k \ket{\alpha^\MF_k}$ in the LLP frame. As shown in Ref.~\cite{Shchadilova2016PRL}, minimization of the variational energy with respect to $\alpha_k^\MF$ leads to the MF solution
\begin{equation}
\alpha_k^\MF = - \frac{V_k^\MF}{\Omega_k^\MF},
\end{equation}
where we defined
\begin{equation}
\Omega_k^\MF=\omega_k+\frac{k^2}{2M} - \frac{k}{M} \l p - P_\ph^\MF \r, \quad V_k^\MF = \beta_\MF V_k.
\end{equation}
In contrast to the Fr\"ohlich case \cite{BeiBing2009,Shashi2014RF}, the scattering amplitudes $V_k^\MF$ are renormalized as well. The two remaining parameters $P_\ph^\MF$ and $\beta^\MF$ are obtained by solving the following set of coupled self-consistency equations,
\begin{equation}
P_\ph^\MF =  (\beta_\MF)^2 \int dk ~ k \frac{V_k^2}{\l \Omega_k^\MF \r^2},
\end{equation}
and
\begin{equation}
\beta_\MF = \left[ 1 + \frac{g_\IB}{2 \pi} \int dk ~ \frac{W_k^2}{\Omega_k^\MF} \right]^{-1}.
\label{eq:selfConsEquations}
\end{equation}

\section{RG approach to Bose polarons\\ in one dimension}
\label{appdx:RGapproach}
In this appendix we extend the MF analysis from Sec.~\ref{subsec:MFtwoPhonon} using a RG approach. We also study the polaron phase diagram of the Bogoliubov polaron model, where phonon-phonon interactions are neglected.

The non-perturbative RG method was developed for the Fr\"ohlich model in Ref.~\cite{Grusdt2015RG}. It was checked by showing excellent agreement with numerically exact diagrammatic Monte Carlo calculations \cite{Vlietinck2015} for the polaron energy. In Ref.~\cite{Grusdt2016RGBEC} we extended this approach to Bose polarons, including two-phonon terms. To benchmark the method also in this case, we compare to our DMC calculations in Fig.~\ref{fig:Intro} (b). In the considered regime of large Bose gas densities $n_0$, corresponding to a small gas parameter $\gamma \lesssim 1$, we find good quantitative agreement.

\subsubsection{RG equations}
\label{subsubsec:RGequations}
Starting from the MF polaron solution $\alpha_k^\MF$ one can rewrite the Hamiltonian $\H_{\rm BF} = \H_F + \H_{\rm 2ph}$ by an effective Hamiltonian $\tilde{\mathcal{H}}$ describing quantum fluctuations around the MF solution. Then the high-energy phonons, with momenta larger than a running UV cut-off $\Lambda$, are eliminated step by step in the RG. This modifies the MF solution $\alpha_k(\Lambda)$ at small energies \cite{Grusdt2016RG}, and gives rise to an RG flow of the coupling constants when $\Lambda$ changes. Using the same notation and following the derivation of Ref.~\cite{Grusdt2016RGBEC} we obtain an effective Hamiltonian
\begin{multline}
\tilde{\mathcal{H}}(\Lambda) = E_0(\Lambda) + \int^\Lambda dk \biggl\{  \Omega_{k} \ad_{k} \a_{k} +  \int^\Lambda  dk' \frac{k k'}{2 \mathcal{M}} : \G_{k} \G_{k'} :  \\
+ \frac{G_+}{2 n_0} : \delta \hat{n}_{k} \delta \hat{n}_{k'}:  +  2 n_0 G_-  :\hat{\vartheta}_{k} \hat{\vartheta}_{k'}:   \biggr\}.
\label{eq:Huniversal}
\end{multline}

A few explanations are in order. First of all, note that the impurity operators $\hat{x}$ and $\hat{p}$ have been eliminated by applying the Lee-Low-Pines transformation \cite{Lee1953} and considering a polaron with vanishing total momentum. The polaron energy $E_0(\Lambda)$, starting at $E_0^\MF$ for $\Lambda=\Lambda_0$, decreases until it reaches the ground state energy of the polaron for $\Lambda \to 0$. The effective phonon frequency in the frame co-moving with the impurity is given by
\begin{equation}
\Omega_{k}(\Lambda) = \omega_k+k^2/2 \mathcal{M}(\Lambda),
\end{equation}
where $\mathcal{M}(\Lambda)$ denotes the renormalized mass of the impurity. Note that $\mathcal{M}(\Lambda \to 0) \approx M_{\rm p}$ can be used as an approximation for the effective polaron mass \cite{Grusdt2016RG}.

The last term in the first line of Eq.~\eqref{eq:Huniversal} describes phonon-phonon interactions induced by the mobile impurity, where the operators $\G_k$ are defined as $\G_k= \ad_{\vec{k}} \a_{\vec{k}} + \alpha_{k} (\ad_{\vec{k}} + \a_{\vec{k}} )$. Note that the MF amplitude is flowing in the RG,
\begin{equation}
\alpha_k(\Lambda)=- \frac{\beta(\Lambda) V_k}{\Omega_k(\Lambda)}.
\label{eq:RGmfAmplitude}
\end{equation}
The second line in Eq.~\eqref{eq:Huniversal} is an alternative formulation of the two-phonon scattering terms $\H_{\rm 2ph}$ in Eq.~\eqref{eq:H2phDef}, with coupling constants $G_\pm(\Lambda)$ running in the RG. We introduced the following pairs of conjugate operators,
\begin{flalign}
\delta \hat{n}_{k} &= \sqrt{n_0} W_k \l \a_{k} + \ad_{-k}  \r, \\
\hat{\vartheta}_{k} &=  \frac{1}{2 i \sqrt{n_0}} W_k^{-1}  \l \a_{k} - \ad_{-k} \r,
\end{flalign}
describing particle number and phase fluctuations of the Bose gas.

The initial conditions for the coupling constants flowing in the RG are given by
\begin{align}
&G_\pm(\Lambda_0) =  \frac{g_\IB}{4 \pi},  &\mathcal{M}(\Lambda_0) = M  \\
&E_0(\Lambda_0) = E_0^\MF, &\beta(\Lambda_0)=\beta_\MF.
\end{align}
In one dimension the RG flow equations read~\cite{Grusdt2016RGBEC},
\begin{flalign}
\partial_\Lambda \mathcal{M} &= - 4  \Lambda^2 \alpha_\Lambda^2 / \Omega_\Lambda \label{eq:RGflowMSymm},\\
\partial_\Lambda G_\pm^{-1} &= - 4  W_\Lambda^{\pm 2} / \Omega_\Lambda, \label{eq:RGflowGpm}
\end{flalign}
and the RG flow of $\beta(\Lambda)$ is described by
\begin{equation}
 \beta(\Lambda) = G_+^{-1}(\Lambda_0) \left[ G_+^{-1}(\Lambda) + 2 \int_{-\Lambda}^\Lambda dp ~ W_p^2 / \Omega_p \right]^{-1}.
\label{eq:g1FormalSolutionSymm}
\end{equation}
The ground state energy can be determined from
\begin{multline}
\partial_\Lambda E_0 = \frac{1}{2} \partial_\Lambda \mathcal{M}^{-1} \int_{-\Lambda}^\Lambda dp ~ \alpha_{p}^2 p^2  + \\
 + \int_{-\Lambda}^\Lambda dp  ~ \frac{2}{\Omega_{\Lambda}} \l W_\Lambda W_p G_+ - W_p^{-1} W_\Lambda^{-1} G_- \r^2.
 \label{eq:RGflowE0}
\end{multline}

In three dimensions~\cite{Grusdt2016RGBEC} all coupling constants converge when the cut-off $\Lambda \to 0$. In that case the RG flows stop when the dispersion relation $\omega_k$ becomes linear for $\Lambda \approx 1/\xi$. Except for $G_-$, this is also true in one dimension. Here the coupling constant $G_-$ always flows to the weak coupling fixed-point in the IR limit,
\begin{equation}
G_-(\Lambda \to 0) =  0^+.
\label{eq:GmZeroFixedPoint}
\end{equation}
To see this, note that we obtain a divergent RG flow $\partial_\Lambda G_-^{-1} \simeq - 1/ \Lambda^2$ when $\Lambda \to 0$.

\subsubsection{Regularization of the IR log-divergence}
\label{subsubsec:IRlogDivRegRG}
The MF polaron energy $E_0^\MF$ from Eq.~\eqref{eq:EoMF} diverges when the IR cut-off $\lambda$ is sent to zero. The reason for this divergence is the unphysical assumption of MF theory that the coupling constant $g_\IB$ is unmodified by quantum fluctuations. Now we show that the RG flow of $G_-$ to the universal weak coupling fixed point $G_-=0$, see Eq.~\eqref{eq:GmZeroFixedPoint}, leads to a regularized polaron energy.

We assume an IR cut-off $\lambda$, where the RG flow is stopped. As discussed around Eq.~\eqref{eq:e0MFlogDiv} the MF energy $E_0^\MF(\lambda)$ has a contribution
\begin{equation}
 E_0^\MF(\lambda) \stackrel{\rm IR}{\simeq}
  \frac{g_\IB}{\pi} \int_\lambda^{\Lambda_0} dk~  \sinh^2 \theta_k  ~ \stackrel{\rm IR}{\simeq} - \frac{g_\IB  \log (\lambda)}{2 \sqrt{2} \pi \xi},
  \label{eq:MFdivergency}
\end{equation}
which diverges logarithmically when $\lambda \to 0$. From the RG we can calculate the polaron energy $E_0^{\rm RG}(\lambda)$ by solving Eq.~\eqref{eq:RGflowE0} for $\Lambda$ flowing from $\Lambda_0$ to $\lambda$, i.e. $E_0^{\rm RG}(\lambda) = E_0(\Lambda = \lambda)$. From the terms in the second line of Eq.~\eqref{eq:RGflowE0} we obtain a contribution
\begin{multline}
\partial_\Lambda E_0 ~  \stackrel{\rm IR}{\simeq}  ~  \frac{4 G_-^2(\Lambda)}{W_\Lambda^2 \Omega_{\Lambda}} \int_\lambda^{\Lambda} dp ~ W_p^{-2}
=  (\partial_\Lambda G_-) \int_\lambda^{\Lambda} dp ~ W_p^{-2} \\
 \stackrel{\rm IR}{\simeq} - (\partial_\Lambda G_-)  \sqrt{2} \xi^{-1} \log (\lambda).
\end{multline}
Integrating this equation yields
\begin{equation}
E_0^{\rm RG}(\lambda) \stackrel{\rm IR}{\simeq} E_0^\MF(\lambda) + \frac{\sqrt{2} \log (\lambda)}{\xi}  \left[ G_-(\Lambda_0) - G_-(\lambda) \right].
\label{eq:cancelLogDiv1}
\end{equation}

From the explicit solution of the RG flow of $G_-^{-1}(\Lambda)$ we obtain the exact expression
\begin{equation}
G_-^{-1}(\lambda) = G_-^{-1}(\Lambda_0) + 4 \int_\lambda^{\Lambda_0} d\Lambda~ \frac{W_\Lambda^{-2}}{\Omega_\Lambda} \stackrel{\rm IR}{\simeq} \lambda^{-1},
\end{equation}
from which it follows that $G_-(\lambda) = \mathcal{O}(1/\lambda)$. I.e. $G_-$ approaches the weak coupling fixed point $G_-=0$ with a power-law in $\lambda$. Because $\log(\lambda) / \lambda \to 0$ for $\lambda \to 0$, the last term in Eq.~\eqref{eq:cancelLogDiv1} is irrelevant in the IR limit.

Finally, combining Eqs.~\eqref{eq:MFdivergency}, \eqref{eq:cancelLogDiv1} and using $G_-(\Lambda_0) = g_\IB / 4 \pi$ we obtain
\begin{equation}
E_0^{\rm RG}(\lambda) \stackrel{\rm IR}{\simeq} \frac{\sqrt{2} \log (\lambda)}{\xi} \left [G_-(\Lambda_0) -  \frac{g_\IB}{4 \pi} \right] =0 \times  \log (\lambda).
\end{equation}
I.e. the two log-divergent terms cancel exactly and the polaron energy obtained from the RG is fully convergent when the IR cut-off $\lambda \to 0$ becomes small.

Before moving on, a comment is in order about the number of phonons in the polaron cloud, which according to MF theory diverges logarithmically with the IR cut-off. In Sec.\ref{subsec:MFtwoPhonon} we argued that this is directly connected to the log-divergence of the MF polaron energy. Now we have proven that the coupling constant $g_\IB$ gives rise to two different couplings $G_\pm$ flowing in the RG, where the renormalization of $G_-$ to zero at low energies regularizes the polaron energy.

We emphasize that the number of phonons in the polaron cloud is still diverging as $\lambda \to 0$, as can be readily checked from analyzing the IR behavior of the renormalized MF amplitudes in Eq.~\eqref{eq:RGmfAmplitude}. As a consequence we expect that the quasiparticle weight $Z = 0$ for $\lambda \to 0$ also within the RG formalism. Therefore we conclude that the orthogonality catastrophe~\cite{Anderson1967} also exists for mobile impurities interacting with 1D quantum gases. As a direct way of detecting this effect for ultracold atoms, Ramsey interferometry can be used as suggested in Ref.~\cite{Knap2012}.

\subsubsection{Polaron phase diagram}
\label{subsubsec:phaseDiag}
Now we analyze the RG flows of the coupling constants more closely and derive the polaron phase diagram. We work in a regime where phonon-phonon interactions $\H_{\rm ph-ph}$ can be neglected. We will show that the phase diagram shares all qualitative features with the 3D case discussed in Ref.~\cite{Grusdt2016RGBEC}.

\emph{Static impurity in a non-interacting Bose gas.--}
Let us start by considering the exactly solvable case of an infinitely heavy impurity, $M = \infty$, localized in the origin. Furthermore we assume that the bosons are non-interacting. For repulsive impurity-boson interactions, $g_\IB > 0$, the ground state corresponds to a wave function where all bosons populate the same single-particle state, forming a repulsive polaron. For arbitrarily weak attraction, $g_\IB < 0$, a bound state $\psi_b(x)$ of bosons to the impurity always exists in one dimension. In this regime the spectrum is unbounded, because $\psi_b(x)$ can be occupied by any integer number of bosons. Note that the MF and RG theories provide a description of the polaron state at finite energy, where no bosons are bound to the impurity~\cite{Grusdt2016RGBEC}. The polaron is meta-stable on the attractive side because it can decay and form a molecule.

In the RG theory the existence of a bound state is indicated by a divergence of the effective interaction strength during the RG flow, $G_\pm(\Lambda) \to \infty$. For the case without boson-boson interactions described above it holds $G_+(\Lambda) = G_-(\Lambda)$. As shown in Eq.~\eqref{eq:GmZeroFixedPoint}, $G_- \to 0^+$ always flows to the repulsive weak coupling fixed point. Because the RG flow starts at $G_\pm(\Lambda_0) = g_\IB/4 \pi$, there always exists a divergence $G_-(\Lambda_c) \to \infty$ at some intermediate $\Lambda_c$ on the attractive side $g_\IB < 0$. As explained in detail in Ref.~\cite{Grusdt2016RGBEC} this is a direct manifestation for the bound state existing at low energies in this regime.

When the mass of the impurity is slowly decreased, we expect the bound states to remain stable because their energy spacings are sizable. Note however that the finite mass of the impurity introduces correlations between the bosons and requires us to solve a full many-body problem.

\emph{Stable repulsive polarons.--}
Now we extend our discussion to finite mass $M < \infty$ and non-vanishing boson-boson interactions $g_{\rm BB} > 0$. The RG flows of $G_\pm(\Lambda)$, which differ in this case, are shown in Fig.~\ref{fig:GpFlow}. On the repulsive side, $g_\IB > 0$, the only qualitative change is that $G_+(0)>0$ saturates at a finite value in the IR limit. In this regime the ground state is a repulsive polaron. In Fig.~\ref{fig:depletion} we show the density profile of the Bose gas around the impurity, and indeed the impurity repels bosons in this regime. For very strong repulsive interactions, the quasi-1D Bose gas is completely depleted around the impurity, reminiscent of the bubble polarons predicted in this regime in Ref.~\cite{blinova2013single} or, equivalently, a dark soliton as described in Sec.~\ref{subsec:GPElimit}.

\emph{Attractive polarons.--}
As discussed above, $G_-$ diverges during the RG flow for arbitrary attractive interactions. On the other hand, the flow of the coupling constant $G_+(\Lambda)$ stops in the IR limit due to finite $g_{\rm BB}>0$. For sufficiently weak attraction,
\begin{equation}
0 > g_\IB > g_{\IB, c}^{\rm RG} = - 2 \pi \biggl( \int_{-\Lambda_0}^{\Lambda_0} d \Lambda~ W_\Lambda^2 / \Omega_\Lambda \biggr)^{-1},  
\end{equation}
this effect is sufficient to prevent $G_+(\Lambda)$ from diverging during the RG and we obtain $G_+(0) < 0$. This corresponds to attractive interactions of the impurity with density fluctuations in the Bose gas and gives rise to an attractive polaron. Compared to the MF result for the critical interaction strength $g_{\IB,c}^{\MF} $, see Eq.~\eqref{eq:gIBcMF}, the RG predicts a transition already somewhat earlier,
\begin{equation}
 0 \geq g_{\IB, c}^{\rm RG} \geq g_{\IB,c}^{\MF},
\end{equation}
as a consequence of the renormalized mass $\mathcal{M} \geq M$.

\begin{figure}[t!]
\centering
\epsfig{file=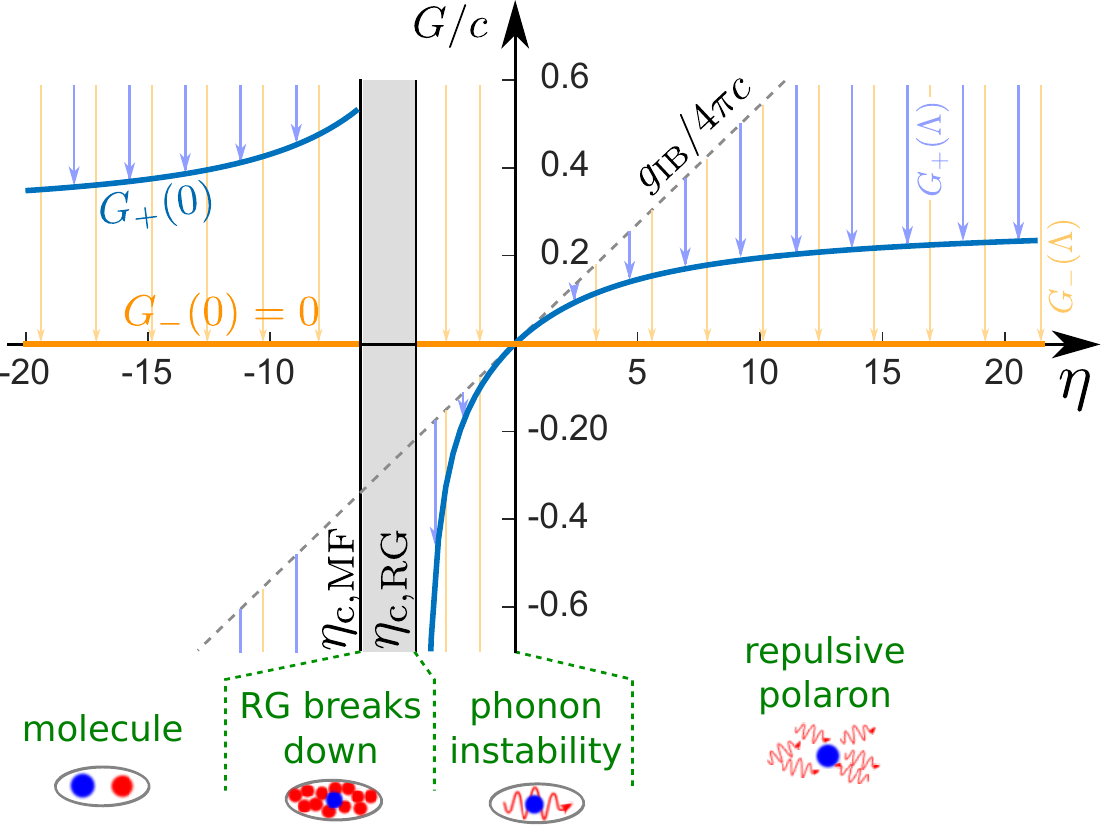, width=0.47\textwidth}
\caption{The RG flows of $G_\pm(\Lambda)$ (blue, yellow) are shown, which start form $g_\IB / 4 \pi$ (indicated by the dashed line). The IR values $G_\pm(0)$ are shown by thick lines; Note that $G_-(0) \equiv 0^+$ and the RG flows have divergencies on the attractive side. The ground states in the different parameter regimes are indicated in the bottom row. In the regime between $\eta_{\rm c, RG} < \eta < \eta_{\rm c, MF}$ the RG breaks down. We used parameters as in the experiment by Catani et al.~\cite{catani2012quantum}.}
\label{fig:GpFlow}
\end{figure}

In Fig.~\ref{fig:GpFlow} the attractive polaron regime, where $0>\eta>\eta_{\rm c,RG} =  g_{\IB, c}^{\rm RG}/g_{\rm BB}$, can be identified. Comparison with Fig.~\ref{fig:1DphaseDiag} shows that, indeed, this is the regime where the polaron energy is negative. The density profile of the Bose gas around the impurity also shows a pronounced peak for this set of parameters, see Fig.~\ref{fig:depletion}.

It is worth emphasizing that \emph{the existence of attractive polarons in one dimension is due solely to non-vanishing boson-boson interactions}. As pointed out before, when $g_{\rm BB} = 0$ the interactions are always repulsive, $G_\pm(0) > 0$, in the long-wavelength limit $\Lambda \to 0$. The effect manifests in the relation $g_{\IB, c}^{\rm RG} = g_{\IB, c}^{\rm MF} = 0$ for $g_{\rm BB} = 0$. In view of this conclusion, the agreement between attractive polaron energies in Fig.~\ref{fig:compHighDensityRep} (a) is remarkable, because it suggests that indeed a meta-stable polaronic eigenstate exists which has negative energy $E_0 < 0$.

The divergence of $G_-$ during the RG in the attractive polaron regime suggests that there exists a mode bound to the impurity at energies below the polaron. In Ref.~\cite{Grusdt2016RGBEC} this effect is discussed in detail and it is shown that the polaron becomes dynamically unstable in this case. Because only $G_-$ is negative, while $G_+$ remains positive, the spectrum of $\H_{\rm BF}$ is continuous and unbounded in the attractive polaron regime \cite{Grusdt2016RGBEC}. In our DMC calculations we fully included phonon-phonon interactions, which are expected to stabilize the attractive polaron \cite{Grusdt2016RGBEC}. Indeed we find a nodeless state at energies corresponding to the attractive polaron, see Fig.~\ref{fig:compHighDensityRep} (a).

\emph{Break-down of the RG.--}
We find that the number of phonons increases dramatically when $g_\IB$ approaches $g_{\IB,c}^{\rm RG}$. In the regime
\begin{equation}
g_{\IB, c}^{\rm RG} > g_\IB > g_{\IB,c}^{\MF},
\end{equation}
the RG breaks down because the MF amplitude diverges at a finite value of $\Lambda$, $\beta(\Lambda) \to \infty$. This is a result of quantum fluctuations of the mobile impurity, because for $M = \infty$ it holds $g_{\IB, c}^{\rm RG}=g_{\IB, c}^{\rm MF}$. Unlike for $G_\pm(\Lambda)$, the divergence of $\beta$ cannot be regularized. As discussion further in Ref.~\cite{Grusdt2016RGBEC} phonon-phonon interactions are required to stop the divergence of the MF amplitude.

\emph{Metastable repulsive polarons.--}
When $g_\IB < g_{\IB,c}^{\MF}$ both coupling constants $G_\pm(\Lambda)$ diverge during the RG flow, see Fig.~\ref{fig:GpFlow}. While they both start out attractive at high energies, they become repulsive in the long-wavelength limit. Therefore the polaron energy $E_0>0$ is positive in this regime, corresponding to a repulsive polaron, see Fig.~\ref{fig:1DphaseDiag}. The repulsive polaron branch is adiabatically connected to the polaronic states realized for repulsive microscopic interactions, similar to the physics of the super-Tonks-Girardeau metastable state \cite{Astrakharchik2005,Haller2009}.

The accumulation of bosons around the impurity is an indicator for molecule formation at $g_\IB < g_{\IB,c}^{\MF}$. Already for $g_\IB \gtrsim g_{\IB,c}^{\RG}$ we find pronounced oscillations in the impurity-boson correlation function, which decay with the distance from the impurity. Indeed, when both coupling constants $G_\pm(\Lambda)$ diverge during the RG, the appearance of a bound state with a discrete energy is expected \cite{Shchadilova2016PRL,Grusdt2016RGBEC}. This state is adiabatically connected to the molecular bound state discussed above at $M=\infty$ and $g_{\rm BB}=0$. In the spectral function it is expected to give rise to a series of peaks separated by the bound state energy~\cite{Shchadilova2016PRL}.

The possibility to decay into molecular states leads to a finite life-time of repulsive polarons when $g_\IB < 0$, a well-known phenomenon close to a Feshbach resonance in three dimensions \cite{Rath2013,Jorgensen2016PRL,Hu2016PRL,Shchadilova2016PRL}. This makes a direct calculation of the polaron energy using DMC difficult, because the polaron is no longer the ground state.

To observe repulsive polarons at $g_\IB < 0$ experimentally, we suggest to study quenches from the strongly repulsive side $g_\IB = +\infty$ to $- \infty$ where the microscopic interactions are attractive. This approach has successfully been used to realize the super-Tonks-Girardeau regime of an interacting 1D gas, see Refs.~\cite{Astrakharchik2005,Haller2009}. We expect that the finite life-time of the repulsive polaron at $g_\IB < 0$ should be observable, for example by using Ramsey interferometry between two spin states which interact differently with the bosons.

\emph{Comparison to the experiment.--}
In Fig.~\ref{fig:1DpolaronMass} we plotted the critical values $\eta_{\rm c, RG}$ and $\eta_{\rm c, MF}$ corresponding to the parameters in Ref.~\cite{catani2012quantum}. For weakly attractive interactions, $0> \eta \gtrsim \eta_{\rm c, RG}$, the measured values for the effective mass are in good agreement with predictions for an attractive polaron. Around $\eta_{\rm c, RG}$ the qualitative behavior of the data changes. The range of parameters $\eta_{\rm c, MF} < \eta < \eta_{\rm c, RG}$ where the RG breaks down and we expect a polaron cloud with many phonons is too narrow to draw any conclusions from the comparison.

\begin{figure}[t!]
\centering
\epsfig{file=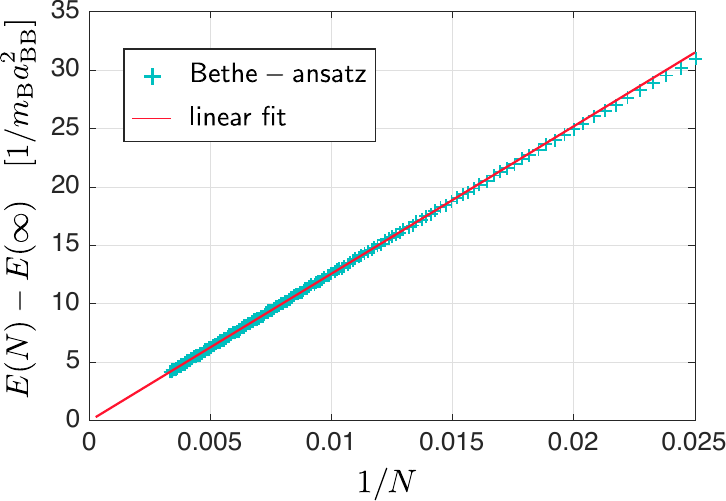,  width=0.43\textwidth} $\quad$
\caption{Finite-size dependence of the ground-state energy $E(N) - E(N=\infty)$ of the bath deep in the Bogoliubov regime, $n_0|a_{\rm BB}| = 144$, and in the absence of the impurity ($\eta=0$). The dependence on the number of atoms in the bath $N$ is obtained from Bethe {\it ansatz} theory \cite{Lieb1963a}. The asymptotic $1/N$ dependence is shown with a solid line obtained as a fit.}
\label{fig:BetheAnsatzFiniteSize}
\end{figure}

\begin{figure}[t!]
\centering
\epsfig{file=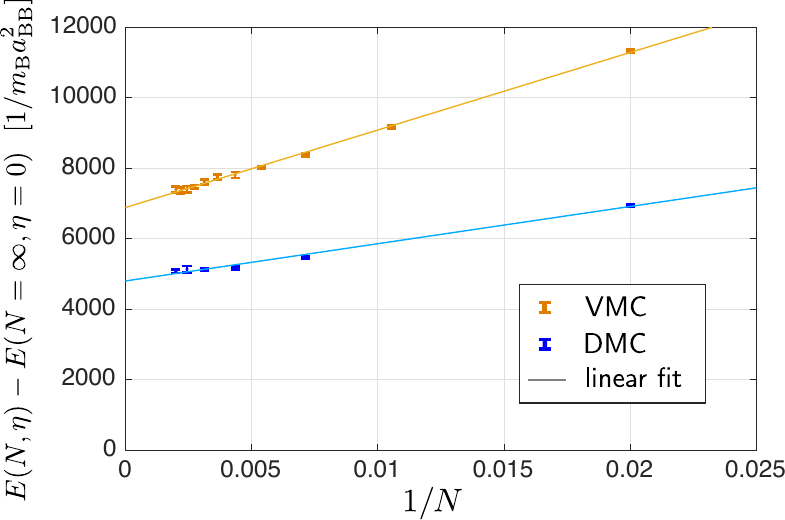, width=0.47\textwidth}
\caption{Finite-size dependence of the polaron energy $E(N,\eta) - E(N=\infty,\eta=0)$ deep in the Bogoliubov regime, $n_0|a_{\rm BB}| = 144$, and for the strongest interaction with the impurity, $\eta=\infty$. The thermodynamic $N\to\infty$ value is obtained from $1/N$ fits, shown with lines.}
\label{fig:MCFiniteSize}
\end{figure}

\section{Finite-size scaling in Monte Carlo calculations}
\label{appdx:finite-size}

The MF and RG theories discussed in the main text predict system properties in the thermodynamic limit of the bath. Instead, quantum Monte Carlo (QMC) simulations are carried out for a finite-size system in a box with periodic boundary conditions. Thus, for making a comparison between different theories it is preferable first to do the extrapolation of the QMC results to the thermodynamic limit. Between two considered densities, corresponding to the Florence experiment \cite{catani2012quantum} and to deep Bogoliubov regime, the latter is expected to have the strongest finite-size effects and we analyze it here.

It is instructive first to study how the energy of the bath depends on the system size in the absence of the impurity. The energy of a single-component Bose gas with $\delta$-pseudopotential interaction can be exactly found using Bethe {\it ansatz} approach \cite{Lieb1963a}. Figure~\ref{fig:BetheAnsatzFiniteSize} shows how the difference of the total energy of the bath and its thermodynamic value $E(N) - E(N=\infty)$ depends on the number of particles $N$. Even if the convergence in the energy per particle $E(N)/N$ has a fast $1/N^2$ dependence for large system sizes, in the total energy the asymptotic dependence is weaker, as can be seen from the $1/N$ fit in Fig.~\ref{fig:BetheAnsatzFiniteSize}.
It should be noted, that the polaron energy is obtained from the total energy $E(N)$, which diverges linearly with number of particles $N$.
This imposes severe requirements for the numerical accuracy goal, especially when large system sizes are used.

For the same high density, $n_0|a_{\rm BB}| = 144$, we now add a finite interaction with the impurity. We consider the extreme case of $\eta=\infty$, corresponding to the strongest interaction. The resulting energy scaling obtained from QMC calculations is reported in Fig.~\ref{fig:MCFiniteSize}. By comparison with the non-interacting case of $\eta=0$ shown in Fig.~\ref{fig:BetheAnsatzFiniteSize} one can see that the effect is greatly enhanced by a finite interaction with the impurity, as can be perceived by contrasting the scales of the vertical axis. At the same time, the asymptotic $1/N$ convergence law is clearly seen.

The polaron energy which we report in the thermodynamic limit in the main part of the paper is obtained from DMC by adjusting a $1/N$ fit to system sizes $N=50,100, 150, 200, 250$.


\end{document}